\documentclass[prd,amsmath,superscriptaddress,,nofootinbib,preprint]{revtex4-2}
\usepackage{amsmath}			
\usepackage{amsfonts}
\usepackage{amssymb}
\usepackage{amsthm}
\usepackage{enumitem}

\usepackage{graphicx}

\begin{document}

\preprint{\today}

\title{A stable Lorentz-breaking model for an antisymmetric two-tensor}

\author{Robertus Potting}
\affiliation{Universidade do Algarve, Faculdade de Ci\^encias e Tecnologia, 8005-139 Faro, Portugal}
\affiliation{CENTRA, Instituto Superior T\'ecnico, Universidade de Lisboa,
Avenida Rovisco Pais, Lisboa, Portugal}

\begin{abstract}
In this work we investigate possible actions for antisymmetric two-tensor field
models subject to constraints that force the field to acquire a nonzero vacuum expectation
value, thereby spontaneously breaking Lorentz invariance.
In order to assure stability, we require that the associate Hamiltonian be bounded from below.
It is shown that this requirement rules out any quadratic action constructed
only from the antisymmetric tensor field.
We then explicitly construct a hybrid model consisting of the antisymmetric two-tensor field
together with a vector field, subject to constraints forcing nonzero expectation values,
that is stable in Minkowski space.
\end{abstract}

\maketitle
\newpage

\section{Introduction}

Among the phenomenological signals that may arise in a theory of quantum gravity
is the possibility that Lorentz invariance be violated. 
One of the possible mechanisms that have been proposed for this
is the spontaneous breaking of Lorentz symmetry
\cite{Kostelecky:1988zi,Kostelecky:1989jw,Kostelecky:1989jp}.
The resulting background values of the fields provide
signatures of Lorentz violation that can be tested experimentally.
A common theoretical framework for their investigation
is provided by the Standard-Model Extension (SME)
\cite{Kostelecky:1994rn,Colladay:1996iz,Colladay:1998fq}.
Most proposals to date in the context of spontaneous Lorentz breaking have concentrated
on the case of a single vector field with a nonderivative potential that depends only
on the squared norm of the vector field,
forcing the vector to acquire a nonzero vacuum expectation value
\cite{Kostelecky:1989jw}.
The vector field can be coupled in General Relativity to the curvature,
leading to preferred-frame effects.
The general class of such theories has been called bumblebee models
\cite{Kostelecky:2003fs,Bluhm:2004ep,Bailey:2006fd,Chkareuli:2007da}.
Among these, a much-studied model, the Einstein-aether theory
\cite{Jacobson:2000xp,Jacobson:2004ts,Deserfest,Carroll:2009en,Withers:2009qg},
considers the most general diffeomorphism-invariant action that is
quadratic in derivatives with fixed norm for the vector field
(which can be imposed by a Lagrange-multiplier potential).
There is an extensive literature on bumblebee models
and their applications in models of gravity
\cite{Gripaios:2004ms,Bertolami:2005bh,Graesser:2005bg,Eling:2006ec,Garfinkle:2007bk,Barausse:2015frm,Casana:2017jkc,Ding:2020kfr},
astrophysics
\cite{Moffat:2002nu,Oost:2018tcv,Khodadi:2020gns,Liang:2022hxd}
and cosmology
\cite{Carroll:2004ai,Bekenstein:2004ne,Zlosnik:2006zu,Tartaglia:2007mh,
Barrow:2012qy}.

An important issue with regard to bumblebee models is that of stability.
A standard way to demonstrate this is to show that the Hamiltonian
is bounded from below.
This requirement severely restricts the possible models.
Namely, the norm of the vector field has to be fixed to a single timelike value,
while the kinetic term has to be of the sigma-model type \cite{Carroll2009,BluhmPotting}.
For any other bumblebee model the Hamiltonian is not bounded from below.
It should be noted that a model with Hamiltonian that is unbounded from below
is not necessarily unstable;
for instance, it might be stabilized by conserved quantities other than the Hamiltonian.
The issue of stability of bumblebee models has also been addressed in
Refs.\ \cite{Clayton:2001vy,Eling:2005zq,Elliott:2005va,Seifert:2007fr,Donnelly:2010qd}.

While the possibility that spontaneous breaking of Lorentz invariance might
occur for tensor fields of order higher than one was proposed early on
\cite{Kostelecky:1994rn},
relatively little studies have appeared in the literature about these.
One exception is Ref.\ \cite{Kostelecky:2009zr},
where a theory of gravity is constructed,
based on the idea that massless gravitons can arise as Goldstone bosons
of broken Lorentz symmetry \cite{Kraus:2002sa,Kostelecky:2005ic}.
Another possibility that has been proposed is that of a Lorentz-breaking
theory of a rank-two antisymmetric tensor field \cite{Altschul2009}.
While a number of studies have appeared applying this idea
(see, e.g., Refs.\ \cite{Seifert:2010uu,Lessa:2019bgi,Assuncao:2019azw,
Maluf:2018jwc,Aashish:2018lhv}),
the study of stability has received little attention.%
\footnote{See Ref.\ \cite{Hernaski:2016dyk} for a study addressing this issue.}
In this work we attempt to fill this gap.
For various reasons it is important to know which models describing the
antisymmetric two-tensor field are stable.
First of all, any studies based on unstable models are,
at the very least, of dubious relevance.
Furthermore, the known existence of any stable models of spontaneous Lorentz breaking
would be a relevant addition to the class of known stable models
that mediate spontaneous Lorentz breaking.

This paper is organized as follows.
In section \ref{sec:bumblebee} we review the (stable) bumblebee model
with sigma-model-type kinetic term.
In section \ref{sec:stability-AS-tensor} we investigate the stability of
the models for the antisymmetric two-tensor field in Minkowski space,
allowing for arbitrary parity-even form of the kinetic term,
quadratic in spacetime derivatives and in the field.
We consider, in alternative, Lagrange multiplier potentials that fix either one
or both of the quadratic invariants that can be built
from the antisymmetric two-tensor field.
As we show, no models with these requisites exist
for which the Hamiltonian is bounded from below.
In order to remedy this, we present in section \ref{sec:hybrid-model}
a hybrid model depending on the antisymmetric two-tensor field together with
a vector field, which has a bounded Hamiltonian.
We analyze some of its properties, including, in section{\ref{sec:linearized-eom},
the linearized equations of motion and a discussion of the Nambu-Goldstone modes.
In section \ref{sec:couplings}, we consider its coupling to gravity and matter.
Finally, in section \ref{sec:conclusions}, we present our conclusions.

In this work we will use natural units with $c = \hbar = 1$,
and the spacetime metric convention $(-+++)$.

\section{Review of the sigma-model bumblebee}
\label{sec:bumblebee}

In this section we will review the stable bumblebee model with
Lagrange-multiplier potential, fixing a timelike expectation value.
 
Consider the Lagrangian density
\begin{align}
\mathcal{L} &= -\frac{1}{2}(\partial_\mu A_\nu)(\partial^\mu A^\nu)
+ \lambda (A_\mu A^\mu + a^2)\nonumber\\
&= \frac{1}{2}\bigl(\dot{A}_i \dot{A}_i - \dot{A}_0^2
- (\partial_i A_j)(\partial_i A_j) + (\partial_i A_0)(\partial_i A_0)\bigl)
+ \lambda(-A_0^2 + A_i A_i + a^2)
\label{L}
\end{align}
Here $A^\mu$ is the bumblebee field, $\lambda$ a Lagrange-multiplier field,
and $a$ a positive constant.
Note that repeated indices are summed over.
The constraint $A_\mu A^\mu + a^2 = 0$ forces $A_\mu$ to take a timelike expectation value,
such that in its rest frame $A^\mu = (\pm a;\vec{0})$.

The canonical momenta are
\begin{align}
\pi_i &= \frac{\delta\mathcal{L}}{\delta\dot{A}_i} = \dot{A}_i\\
\pi_0 &= \frac{\delta\mathcal{L}}{\delta\dot{A}_0} = -\dot{A}_0\\
\pi_\lambda &= \frac{\delta\mathcal{L}}{\delta\dot{\lambda}}  \approx 0
\label{pi-lambda}
\end{align}
so there is one primary constraint.
Note that the ``$\approx$'' sign indicates weak equality in Dirac's
language, that is, equality on the constraint surface in phase space.
The canonical Hamiltonian becomes
\begin{align}
\mathcal{H}_c &= \pi_i\dot{A}_i + \pi_0\dot{A}_0 - \mathcal{L}\nonumber\\
&= \frac{1}{2}\bigl(\dot{A}_i \dot{A}_i - \dot{A}_0^2
+ (\partial_i A_j)(\partial_i A_j) - (\partial_i A_0)(\partial_i A_0)\bigl)
+ \lambda(A_0^2 - A_i A_i - a^2)
\label{H_c}
\end{align}
Following the Dirac procedure, we define the primary Hamiltonian $\mathcal{H}_p$
by adding to \eqref{H_c} the primary constraint multiplied by a Lagrange multiplier:
\begin{equation}
\mathcal{H}_p = \mathcal{H}_c + \mu_\lambda \pi_\lambda
\end{equation}
which defines time evolution in the extended phase space.
Next we impose that the constraint \eqref{pi-lambda} is conserved in time,
i.e., it must have zero Poisson bracket with $\mathcal{H}_p$.
This yields the secondary constraint
\begin{equation}
\pi_\lambda^{(2)} = A_0^2 - A_i A_i - a^2 \approx 0
\label{pi-lambda-2}
\end{equation}
Continuing this process (that is, imposing that the constraint \eqref{pi-lambda-2}
be conserved in time as well can be shown to yield two more constraints in
phase space \cite{BluhmPotting}.
They are all second class, that is, the matrix of their mutual Poisson brackets
is nondegenerate on the constraint surface,
and the number of dynamical degrees of freedom is $5 - 4/2 = 3$.

It is convenient to take them to correspond to the spatial components $A_i$.
Then $A_0$ is fixed by the constraint \eqref{pi-lambda-2}:
\begin{equation}
A_0 = \pm \sqrt{A_iA_i + a^2}
\label{A_0}
\end{equation}
and is thus not an independent dynamical degree of freedom.
As a consequence, it is eliminated as a potential ghost degree of freedom
(note that its time derivative appears with the wrong sign in Eqs.\ \eqref{L}
and \eqref{H_c}).
It then follows from \eqref{A_0} that 
\begin{equation}
\dot{A}_0 = \frac{A_i\dot{A}_i}{A_0}
\end{equation}
so that
\begin{equation}
\dot{A}_0^2 = \frac{(A_i\dot{A}_i)(A_j\dot{A}_j)}{A_0^2}
\le \frac{(A_iA_i)(\dot{A}_j\dot{A}_j)}{A_0^2}
\le \dot{A_i}\dot{A}_i
\end{equation}
where in the first inequality we used Schwarz's inequality
and in the second one constraint \eqref{pi-lambda-2}.
In the same way, it follows that
\begin{equation}
(\partial_i A_0)(\partial_i A_0) \le (\partial_i A_j)(\partial_i A_j)\>.
\end{equation}
Consequently, we find that
\begin{equation}
\mathcal{H}_p \approx \frac{1}{2}\bigl(\dot{A}_i \dot{A}_i - \dot{A}_0^2
+ (\partial_i A_j)(\partial_i A_j) - (\partial_i A_0)(\partial_i A_0)\bigl)
\ge 0
\end{equation}
and we can conclude that the model is stable
\cite{Carroll2009,BluhmPotting}.

It is instructive to consider the equations of motion following from
the Lagrangian \eqref{L}:
\begin{align}
\label{eq1}
\square A_\mu + 2\lambda A_\mu &= 0 \\
A_\mu A^\mu &= -a^2
\label{eq2}
\end{align}
where $\square = \partial_\lambda\partial^\lambda$.
Multiplying Eq.\ \eqref{eq1} by $A^\mu$ and using the constraint
\eqref{eq2} yields
\begin{equation}
\lambda = \frac{1}{2a^2}A^\mu \square A_\mu\>.
\label{lambda}
\end{equation}
Substituting Eq.\ \eqref{lambda} back into \eqref{eq1} yields
the factored expression
\begin{equation}
\left(\delta^\nu_\mu + \frac{1}{a^2}A_\mu A^\nu \right)\square A_\nu = 0
\label{bumblebee-factored}
\end{equation}
in which the factor in front projects onto fluctuations $\delta A_\mu$
that are transverse to $A_\mu$, satisfying $A^\mu \delta A_\mu = 0$. 
This shows clearly that there are three dynamical massless degrees of freedom.

It can be shown that the choice of other kinetic terms than the
``sigma-model-type'' Lagrangian \eqref{L} always leads
to Hamiltonians that are not bounded from below \cite{Carroll2009,BluhmPotting}.
For the case of the Maxwell kinetic term it turns out that there
are actually no linear modes that destabilize this model \cite{Carroll2009}.
Also, the latter can be shown to be stable on the restricted phase space with
$\lambda = 0$ \cite{BluhmPotting}.
In that case, the Lagrange multiplier term no longer contributes any
restoring force in the equations of motion for the vector field
and, as a consequence,
the dynamics of the model coincides with that of pure Maxwell theory.
However, this is no longer the case whenever the Lagrange multiplier becomes
nonzero at any point in space, which may happen if the vector field
is coupled to external fields.
In those situations instabilities can be expected to set in.
For instance, it is known that the Maxwell kinetic term in models
with spontaneous Lorentz breaking can lead to the spontaneous formation
of shock waves \cite{Clayton:2001vy}.

\section{Attempt to formulate stable model with the antisymmetric two-tensor field}
\label{sec:stability-AS-tensor}

Next consider the antisymmetric two-tensor field $B_{\mu\nu} = -B_{\nu\mu}$.
It was considered first in the context of a Lorentz-violating model
in \cite{Altschul2009}, where it was pointed out that the most general
nonderivative potential is of the form $V(Y_1,Y_2)$,
where $Y_1 = B_{\mu\nu}B^{\mu\nu}$ and
$Y_2 = \epsilon^{\mu\nu\rho\sigma}B_{\mu\nu}B_{\rho\sigma}$
are the only two functionally independent (pseudo)scalar combinations
that can be built from $B_{\mu\nu}$.
Consequently, for suitable potential,
$Y_1$ and $Y_2$ can independently acquire a vacuum expectation value.
If either is nonzero,
this implies that Lorentz invariance is broken spontaneously.
The various ways in which this can happen, as well as the dynamics of
these models, are discussed at length in \cite{Altschul2009}.

However, one issue that remained unexplored in \cite{Altschul2009}
is that of stability.
Here we will investigate the stability of models with the most general possible
kinetic term that is quadratic in $B_{\mu\nu}$.
For the potential we will first consider the Lagrange-multiplier potential
\begin{equation}
V_{\lambda_1\lambda_2} = \lambda_1(B_{\mu\nu}B^{\mu\nu} -  y_1) + \lambda_2(\epsilon^{\mu\nu\rho\sigma}B_{\mu\nu}B_{\rho\sigma} - y_2)
\label{V-lambda1-lambda2}
\end{equation}
which fixes the values of $Y_1$ and $Y_2$,
not allowing any fluctuations from their vacuum value.
In alternative, we will also consider the potential
\begin{equation}
V_{\lambda_1} = \lambda_1(B_{\mu\nu}B^{\mu\nu} -  y_1)
\label{V-lambda1}
\end{equation}
which fixes only the value of $Y_1$, leaving $Y_2$ free.
It would be interesting to consider also the case of a smooth potential.
However, this is beyond the scope of the current work.

The most general kinetic term that is quadratic in $B_{\mu\nu}$ and parity even is
\begin{equation}
\label{L_B-general}
\tfrac{1}{2}\sigma_1\bigl(\nabla_\lambda B^{\mu\nu}\bigr)\bigl(\nabla^\lambda B_{\mu\nu}\bigr)
+ \tfrac{1}{2}\sigma_2\bigl(\nabla_\lambda B^{\mu\nu}\bigr)\bigl(\nabla_\mu B^\lambda{}_\nu\bigr)
+ \tfrac{1}{2}\sigma_3\bigl(\nabla_\mu B^{\mu\nu}\bigr)\bigl(\nabla_\lambda B^\lambda{}_\nu\bigr)\>.
\end{equation}
Passing to flat space, and performing a partial integration of the second term,
upon inclusion of the potential terms,
the Lagrangian density takes either one of the alternative forms
\begin{align}
\label{L_flat-space_lambda1-lambda2}
\mathcal{L}_{\lambda_1\lambda_2} &= \tfrac{1}{2}\sigma_1\bigl(\partial_\lambda B^{\mu\nu}\bigr)\bigl(\partial^\lambda B_{\mu\nu}\bigr)
+ \tfrac{1}{2}\sigma_{23}\bigl(\partial_\mu B^{\mu\nu}\bigr)\bigl(\partial_\lambda B^\lambda{}_\nu\bigr) - V_{\lambda_1\lambda_2}\\
\mathcal{L}_{\lambda_1} &= \tfrac{1}{2}\sigma_1\bigl(\partial_\lambda B^{\mu\nu}\bigr)\bigl(\partial^\lambda B_{\mu\nu}\bigr)
+ \tfrac{1}{2}\sigma_{23}\bigl(\partial_\mu B^{\mu\nu}\bigr)\bigl(\partial_\lambda B^\lambda{}_\nu\bigr) - V_{\lambda_1}
\label{L_flat-space_lambda1}
\end{align}
where $\sigma_{23} = \sigma_2 + \sigma_3$.

In order to assess the stability of Lagrangian \eqref{L_flat-space_lambda1-lambda2},
we will pass to phase space, carry out a careful analysis of the
constraint structure, determine the Hamiltonian and verify if it is bounded from below.
As it turns out, there are some special values for the coefficients $\sigma_1$
and $\sigma_{23}$ for which the constraint structure is different from the general case.
Therefore, we will distinguish the following three cases:
\begin{enumerate}[label={\Alph*.},noitemsep]
\item $\sigma_1 \ne 0$, $\sigma_{123}\ne 0$;
\item $\sigma_1 \ne 0$, $\sigma_{123} = 0$;
\item $\sigma_1 = 0$, $\sigma_{123} \ne 0$.
\end{enumerate}
Here we defined $\sigma_{123} = \sigma_1 + \frac{1}{2}\sigma_{23}$.

\subsection{Case $\sigma_1 \ne 0$, $\sigma_{123}\ne 0$}
\label{subsec: general_case}

We start with analyzing the canonical structure of Lagrangian \eqref{L_flat-space_lambda1-lambda2}
for the general case, such that neither $\sigma_1$ nor the combination
$\sigma_{123}$ vanishes.
First we introduce the canonical momenta
\begin{align}
\label{pi_0i}
\pi_{0i} &= \frac{\delta\mathcal{L}_{\lambda_1\lambda_2}}{\delta\dot{B}_{0i}}
= 2\sigma_{123}\dot{B}_{0i} + \sigma_{23}\partial_j B_{ij} \\
\pi_{ij} &= \frac{\delta\mathcal{L}_{\lambda_1\lambda_2}}{\delta\dot{B}_{ij}}
= -\sigma_1 \dot{B}_{ij}
\label{pi_ij}
\end{align} 
where $\sigma_{123} = \sigma_1 + \frac{1}{2}\sigma_{23}$,
and the overdot stands for derivative with respect to time.
The momenta associated with the Lagrange multipliers are identically zero,
so that they will give rise, in Dirac's terminology, to the primary constraints
\begin{align}
\label{phi_1}
\phi_1 = \pi_{\lambda_1} &\approx 0 \\
\phi_2 = \pi_{\lambda_2} &\approx 0
\end{align}
The canonical Hamiltonian can be written as
\begin{align}
\mathcal{H}_c &= \pi_{0i}\dot{B}_{0i} + \pi_{ij}\dot{B}_{ij} - \mathcal{L}_{\lambda_1\lambda_2}\nonumber\\
&= \frac{1}{4\sigma_{123}}\bigl(\pi_{0i} + \sigma_{23}\partial_j B_{ij}\bigr)
  \bigl(\pi_{0i} + \sigma_{23}\partial_k B_{ik}\bigr)
- \frac{1}{2\sigma_1}\pi_{ij} \pi_{ij}
+ \sigma_1\bigl(\partial_i B_{0j}\bigr)\bigl(\partial_i B_{0j}\bigr)
\nonumber\\
&\qquad{}
+ \tfrac{1}{2}\sigma_{23}\bigl(\partial_{i}B_{0i}\bigr)^2
- \tfrac{1}{2}\sigma_1\bigl(\partial_i B_{jk}\bigr)\bigl(\partial_i B_{jk}\bigr)
- \tfrac{1}{2}\sigma_{23}\bigl(\partial_j B_{ij}\bigr)\bigl(\partial_k B_{ik}\bigr)
+ V_{\lambda_1\lambda_2}\>.
\label{H_c}
\end{align}
Following Dirac's procedure, one now defines the total Hamiltonian by
adding to $\mathcal{H}_c$ a linear combination of the primary constraints:
\begin{equation}
\mathcal{H}_T = \mathcal{H}_c + \mu_1\phi_1 + \mu_2\phi_2\>.
\label{H_T}
\end{equation}
Here $\mu_1$ and $\mu_2$ are, as yet, undetermined parameters.
Next one imposes that the primary constraints be conserved in time, where
time evolution is to be generated by the total Hamiltonian $H_T = \int \mathcal{H}_T\,d^3x$.
We thus obtain the secondary constraints
\begin{align}
\label{phi^2_1}
\phi^{(2)}_1 &= \{\phi_1,H_T\} = -2B_{0i}B_{0i} + B_{ij}B_{ij} - y_1 \approx 0\\
\phi^{(2)}_2 &= \{\phi_2,H_T\} = -4\epsilon_{0ijk} B_{0i} B_{jk} - y_2 \approx 0
\label{phi^2_2}
\end{align}
which are, of course, the constraints enforced by the Lagrange multipliers
$\lambda_1$ and $\lambda_2$.
Demanding that $\phi^{(2)}_1$ and $\phi^{(2)}_2$ be conserved in time yields
two more constraints:
\begin{align}
\label{phi^3_1}
\phi^{(3)}_1 &= \{\phi^{(2)}_1,H_T\} = -4B_{0i}\dot{B}_{0i} + 2B_{ij}\dot{B}_{ij}\nonumber\\
&= -\frac{2}{\sigma_{123}}B_{0i}\bigl(\pi_{0i} + \sigma_{23}\partial_j B_{ij}\bigr)
- \frac{2}{\sigma_1}B_{ij}\pi_{ij}
\approx 0\\
\label{phi^3_2}
\phi^{(3)}_2 &= \{\phi^{(2)}_2,H_T\} = -4\epsilon_{0ijk}(\dot{B}_{0i}B_{jk} + B_{0i}\dot{B}_{jk}\nonumber\\
&= -4\epsilon_{0ijk}\left(\frac{1}{2\sigma_{123}}\bigl(\pi_{0i} + \sigma_{23}\partial_l B_{il}\bigr)B_{jk}
+ \frac{1}{\sigma_1}B_{0i}\pi_{jk}\right)
\approx 0
\end{align}
Iterating this process yields yet one more set of secondary constraints:
\begin{align}
\label{phi^4_1}
\phi^{(4)}_1 &= \{\phi^{(3)}_1,H_T\} \nonumber\\
&=-\frac{1}{2\sigma_{123}^2}\bigl(\pi_{0i} + \sigma_{23}\partial_j B_{ij}\bigr)
\bigl(\pi_{0i} + \sigma_{23}\partial_k B_{ik}\bigr)
+ \frac{1}{\sigma_1^2}\pi_{ij}\pi_{ij}\nonumber\\
&\qquad{}
+ \frac{B_{0i}}{\sigma_{123}}\left(-2\sigma_1\partial^2B_{0i}-\sigma_{23}\partial_i\partial_j B_{0j}
-4\lambda_1 B_{0i} - 4\lambda_2\epsilon_{0ijk}B_{jk} + \frac{\sigma_{23}}{\sigma_1}\partial_j\pi_{ij}\right)\nonumber\\
&\qquad{}
+ \frac{B_{ij}}{\sigma_1}\left(\sigma_1\partial^2 B_{ij} + \sigma_{23}\partial_j\partial_k B_{ik}
- \frac{\sigma_{23}}{2\sigma_{123}}\partial_j(\pi_{0i} + \sigma_{23}\partial_k B_{ik})
+ 2\lambda_1 B_{ij} - 4\lambda_2\epsilon_{0ijk}B_{0k}\right)\nonumber\\
&\approx 0\\
\label{phi^4_2}
\phi^{(4)}_2 &= -\{\phi^{(3)}_2,H_T\} \nonumber\\
&= 4\epsilon_{0ijk}\biggl[\frac{1}{2\sigma_{123}}\left(2\sigma_1\partial^2 B_{0i} + \sigma_{23}\partial_i\partial_j B_{0j} + \frac{\sigma_{23}}{\sigma_1}
\partial_l\pi_{il}\right)B_{jk}\nonumber\\
&\qquad\qquad{}
+ \frac{1}{\sigma_1}\biggl(-\frac{1}{2\sigma_{123}}(\pi_{0i} + \sigma_{23}\partial_l B_{il})\pi_{jk}\nonumber\\
&\qquad\qquad\qquad{}
+ B_{0i}\Bigl(\sigma_1\partial^2B_{jk} + \sigma_{23}\partial_k\partial_l B_{jl}
- \frac{\sigma_{23}}{2\sigma_{123}}\partial_k(\pi_{0j} 
+ \sigma_{23}\partial_l B_{jl})\nonumber\\
&\qquad\qquad\qquad\qquad{}
 + 2\lambda_1 B_{jk} - 4\lambda_2\epsilon_{0ljk}B_{0l}\Bigr)\biggr)\biggr]
\approx 0
\end{align}
For the time evolutions of $\phi^{(4)}_1$ and $\phi^{(4)}_2$ we find expressions
of the form
\begin{align}
\label{dot-phi^4_1}
\{\phi^{(4)}_1,H_T\} &=
-2\mu_1\left(-\frac{2}{\sigma_{123}}B_{0i}B_{0i} + \frac{1}{\sigma_1}B_{ij}B_{ij}\right)
+ 4\mu_2\left(\frac{1}{\sigma_{123}} + \frac{1}{\sigma_1}\right)
\epsilon_{0ijk}B_{0i}B_{jk} + \ldots \\
\{\phi^{(4)}_2,H_T\} &= 8\mu_1\left(\frac{1}{\sigma_{123}} + \frac{1}{\sigma_1}\right)\epsilon_{0ijk}B_{0i}B_{jk}
+ 16\mu_2\left(-\frac{2}{\sigma_{123}}B_{0i}B_{0i} + \frac{1}{\sigma_1}B_{ij}B_{ij}\right) + \ldots
\label{dot-phi^4_2}
\end{align}
where the ellipses denote terms that do not involve the coefficients $\mu_i$.
We can impose that expressions \eqref{dot-phi^4_1} and \eqref{dot-phi^4_2}
vanish strongly by solving them for $\mu_1$ and $\mu_2$.
Consequently, there are no further secondary constraints.

The phase space has $2 \times (6 + 2) = 16$ local degrees of freedom
(corresponding to the 6 independent components of $B_{\mu\nu}$
and the Lagrange multipliers, together with their canonical momenta)
while there are 8 remaining constraints.
It can be checked that they are second class.
Therefore, altogether, the system defined by Lagrangian \eqref{L_flat-space_lambda1-lambda2}
has 8 degrees of freedom in phase space.
\\

Next let us see if there is a subset of values of the coefficients $\sigma_i$
for which the Hamiltonian is bounded from below.
First of all,
it follows from Eq.\ \eqref{H_c} that, in order to avoid ghost instabilities,
one needs to impose the restrictions
\begin{equation}
\sigma_{123} > 0 \qquad\text{and}\qquad \sigma_1 < 0\>.
\label{no-ghosts}
\end{equation}
In terms of the vectors $H_i = \tfrac{1}{2}\epsilon_{0ijk}B_{jk}$ and
$E_i = B_{0i}$,
the constraints \eqref{phi^2_1} and \eqref{phi^2_2} imply that
\begin{equation}
|\vec{H}\,|^2 - |\vec{E}\,|^2 = -\tfrac{1}{2}y_1 
\qquad \text{and}\qquad
\vec{H}\cdot\vec{E}  = -\tfrac{1}{2}y_2 \>.
\label{constraints_AC}
\end{equation}
Now consider any configuration with cartesian coordinates
\begin{equation}
\vec{H} = (H_\perp\cos\alpha\>,\>H_\perp\sin\alpha\>,\>H_\|)\>,\qquad
\vec{E} = (0\>,\>0\>,\>E)
\end{equation}
with constants $H_\|$, $H_\perp$ and $E$ satisfying
\begin{equation}
H_\perp^2 + H_\|^2 - E^2 = -\tfrac{1}{2}y_1 
\qquad \text{and}\qquad
H_\|E  = -\tfrac{1}{2}y_2 \>.
\label{constraints_AC2}
\end{equation}
and $\alpha(\vec{r})$ some function of the spatial coordinates.
The conditions \eqref{constraints_AC2} can be readily solved
for $H_\|$ and $E$ for any values of $y_1$, $y_2$ and $H_\perp$.
It is easy to check that, if $H_\perp$ is taken to be nonzero and $H_\perp^2 > -\frac{1}{2}y_1$,
it follows from Eqs.\ \eqref{constraints_AC2} that $E \ne 0$ even if $y_2 = 0$. 
Moreover, let us take $\pi_{ij} = \pi_{0i} + \sigma_{23}\partial_k B_{ik} = 0$,
solving constraints \eqref{phi^3_1} and \eqref{phi^3_2}.
Constraints \eqref{phi^4_1} and \eqref{phi^4_2} amount to conditions on the Lagrange
multipliers $\lambda_1$ and $\lambda_2$ of the form
\begin{align}
\label{eq1-lambda-general}
(H_\|^2 + H_\perp^2)\lambda_1 - 2H_\| E\lambda_2 + \ldots &=0 \\
H_\| E \lambda_1 - 4E^2 \lambda_2 + \ldots &= 0
\label{eq2-lambda-general}
\end{align}
where the ellipses correspond to expressions independent of $\lambda_1$ and $\lambda_2$.
The system \eqref{eq1-lambda-general}-\eqref{eq2-lambda-general} can be readily solved
for the Lagrange multipliers in terms of $\vec{H}$ and $\vec{E}$.
This shows the field configuration defined by conditions
\eqref{constraints_AC}--\eqref{constraints_AC2} is compatible with the constraints.

The Hamiltonian density now becomes
\begin{align}
\mathcal{H}_c &\approx
-\sigma_{123}(\partial_i H_j)(\partial_i H_j)
+ \tfrac12 \sigma_{23}(\partial_i H_i)^2 \nonumber\\
&= -\sigma_{123} H_\perp^2\bigl((\partial_x\alpha)^2 + (\partial_y\alpha)^2 + (\partial_z\alpha)^2\bigr)
+ \tfrac12 \sigma_{23} H_\perp^2\bigl(-(\sin\alpha)\partial_x\alpha + (\cos\alpha)\partial_y\alpha\bigr)^2\>.
\label{H_c_general-case}
\end{align}
As the two terms in the last expression have opposite signs
(note that $\tfrac12 \sigma_{23} > \sigma_{123} > 0$),
it is not immediately obvious if $\mathcal{H}_c$ is positive definite,
or whether negative values are possible.
In fact,
it is easy to demonstrate that $\alpha(\vec{r})$
can be chosen such that $\mathcal{H}_c$ is nonzero on a local region of space
such that the energy $\int d^3x\,\mathcal{H}_c$ takes any finite negative value. 
For example, let us take 
\begin{equation}
\alpha(\vec{r}) = \epsilon \exp\left[-\bigl(\kappa_\rho(x^2 + y^2) + \kappa_z z^2\bigr)/2\right]
\end{equation}
with positive constants $\epsilon$, $\kappa_\rho$ and $\kappa_z$.
Substitution in Eq.\ \eqref{H_c_general-case} yields
\begin{equation}
\mathcal{H}_c \approx H_\perp^2\epsilon^2
\left[-\sigma_{123}(\kappa_\rho^2\rho^2 + \kappa_z^2z^2) + \tfrac12\sigma_{23} \kappa_\rho^2\rho^2\sin^2\bigl(\alpha(\vec{r}) - \phi\bigr)\right]e^{-\kappa_\rho\rho^2 - \kappa_z z^2}
\label{H_c_general-case2}
\end{equation}
where we introduced cilindrical coordinates, with $x = \rho\cos\phi$ and $y = \rho\sin\phi$.
Expression \eqref{H_c_general-case2} can be readily integrated over space, yielding
for the Hamiltonian (i.e., the energy)
\begin{equation}
\int d^3x \mathcal{H}_c = \frac{\pi^{3/2}H_\perp^2\epsilon^2}{2\kappa_\rho\kappa_z^{1/2}}
\bigl((-2\sigma_{123} + \tfrac12 \sigma_{23})\kappa_\rho - \sigma_{123}\kappa_z\bigr)\>.
\label{energy-general}
\end{equation}
While the sign of the first term in this expression can be either positive or negative,
the last term is negative.
Consequently, we can choose the values of $\epsilon$, $\kappa_\rho$ and $\kappa_z$
such that the energy takes any negative value.
\\

Next, we consider  Lagrangian \eqref{L_flat-space_lambda1}.
There is now only the single primary constraint $\phi_1$ given by expression \eqref{phi_1},
so in the canonical Hamiltonian \eqref{H_c}, the potential term is substituted by $V_{\lambda_1}$,
while the total Hamiltonian \eqref{H_T} only includes the term with the parameter $\mu_1$. 
Applying Dirac's procedure yields the secondary constraints $\phi_1^{(i)}$ ($i=2,3,4$)
given by expressions \eqref{phi^2_1}, \eqref{phi^3_1} and \eqref{phi^4_1},
with $\lambda_2$ set to zero.
The time evolution of $\phi_1^{(4)}$ is of the form of Eq.\ \eqref{dot-phi^4_1},
with $\mu_2$ set to zero.
This condition can be immediately solved for $\mu_1$, so there are no further
constraints. 
The remaining four constraints can be shown to be second class, and thus
it follows that for this case there are $2\times(6 + 1) - 4 = 10$ degrees of freedom
in phase space.

The destabilizing configuration proposed above for the case of two constraints can be applied
to this case as well.
\\

We therefore conclude that, of both $\sigma_1$ and $\sigma_{123}$ are nonzero,
the Hamiltonian is always unbounded from below.

\subsection{Case $\sigma_1 \ne 0$, $\sigma_{123} = 0$}
\label{subsec:sigma123=0}

In this case Lagrangian \eqref{L_flat-space_lambda1-lambda2} becomes
\begin{equation}
\mathcal{L}_{\lambda_1\lambda_2} = \tfrac{1}{2}\sigma_1\left[\bigl(\partial_\lambda B^{\mu\nu}\bigr)\bigl(\partial^\lambda B_{\mu\nu}\bigr)
- 2\bigl(\partial_\mu B^{\mu\nu}\bigr)\bigl(\partial_\lambda B^\lambda{}_\nu\bigr)\right] - V_{\lambda_1\lambda_2}\>.
\end{equation}
This Lagrangian can be cast in a more convenient form by performing a partial integration in the action
on the second term, yielding
\begin{equation}
\mathcal{L}_{\lambda_1\lambda_2} = \tfrac{1}{6}\sigma_1 H_{\lambda\mu\nu} H^{\lambda\mu\nu} - V_{\lambda_1\lambda_2}
\label{L_flat-space_sigma123=0}
\end{equation}
where $H_{\lambda\mu\nu} = \partial_\lambda B_{\mu\nu} + \partial_\mu B_{\nu\lambda} + \partial_\nu B_{\lambda\mu}$
is the field strength associated with $B_{\mu\nu}$, invariant under the gauge transformation
$B_{\mu\nu} \to B_{\mu\nu} + \partial_\mu\Lambda_\nu - \partial_\nu \Lambda_\mu$,
for arbitrary gauge parameter $\Lambda_\mu$.
It is straightforward to demonstrate that, in the absence of the potential term $V_{\lambda_1\lambda_2}$,
the Hamiltonian associated to Lagrangian \eqref{L_flat-space_sigma123=0} is non-negative definite for $\sigma_1 < 0$,
defining a stable gauge-invariant system.
However, we will see that, in the presence of the potential, this is no longer the case.

In order to derive the Hamiltonian and the associated constraints,
we write Lagrangian \eqref{L_flat-space_sigma123=0} as
\begin{align}
\mathcal{L}_{\lambda_1\lambda_2} &= -\tfrac12 \sigma_1\bigl[H_{0ij}H_{0ij} - \tfrac13 H_{ijk}H_{ijk}\bigr] - V_{\lambda_1\lambda_2}\nonumber\\
&= -\tfrac12 \sigma_1\bigl[(\dot B_{ij} + 2\partial_{[i} B_{j]0})(\dot B_{ij} + 2\partial_{[i} B_{j]0})
-(\partial_i B_{jk})(\partial_i B_{jk}) + 2(\partial_j B_{ij})(\partial_k B_{ik})\bigr] - V_{\lambda_1\lambda_2}
\end{align}
where we performed a partial integration in the last term.
For the momentum conjugate to $B_{ij}$ we find
\begin{equation}
\pi_{ij} = -\sigma_1(\dot B_{ij} + 2\partial_{[i} B_{j]0})
\end{equation}
while the momenta conjugate to $B_{0i}$ and the Lagrange multipliers vanish identically,
leading to the primary constraints
\begin{align}
\label{phi_1_B}
\phi_1 = \pi_{\lambda_1} &\approx 0 \\
\phi_2 = \pi_{\lambda_2} &\approx 0 \\
\phi_{0i} = \pi_{0i} &\approx 0 \qquad (i = 1,2,3)
\label{phi_0i}
\end{align}
We obtain for the canonical Hamiltonian
\begin{align}
\mathcal{H}_c &= \pi_{0i}\dot{B}_{0i} - \mathcal{L}_{\lambda_1\lambda_2}\nonumber\\
&= -\frac{1}{2\sigma_1} \pi_{ij}\pi_{ij} + 2\pi_{ij}\partial_i B_{0j}
- \frac{\sigma_1}{2}(\partial_i B_{jk})(\partial_i B_{jk})
+ \sigma_1(\partial_i B_{ij})(\partial_k B_{ik})
+ V_{\lambda_1\lambda_2}\>,
\label{H_c_sigma123=0}
\end{align}
while total Hamiltonian becomes:
\begin{equation}
\mathcal{H}_T = \mathcal{H}_c + \mu_1\phi_1 + \mu_2\phi_2 + \mu_{0i}\phi_{0i}\>,
\label{H_T-caseB}
\end{equation}
where $\mu_1$ and $\mu_2$ and $\mu_{0i}$ are, as yet, undetermined parameters.

Just like in the previous case, we now follow Dirac's procedure and determine the secondary constraints,
by imposing that the primary constraints be conserved in time, where
time evolution is generated by the total Hamiltonian $H_T = \int \mathcal{H}_T\,d^3x$.
This yields the secondary constraints \eqref{phi^2_1} and \eqref{phi^2_2}, together with
\begin{equation}
\phi_{0i}^{(2)} = 2\partial_j\pi_{ji} - 4\lambda_1 B_{0i} - 4\lambda_2\epsilon_{0ijk}B_{jk} \approx 0
\label{phi^2_0i}
\end{equation}
Imposing that the secondary constraints be conserved in time gives the (tertiary) constraints
\begin{align}
\phi_1^{(3)} &= -4\mu_i B_{0i} - \frac{2}{\sigma_1}B_{jk}(\pi_{jk} + 2\sigma_1 \partial_{[i}B_{k]0})\approx 0\\
\phi_2^{(3)} &= \mu_i\epsilon_{0ijk}B_{jk} - \frac{1}{\sigma_1}\epsilon_{0ijk}B_{0i}(\pi_{jk} + 2\sigma_1 \partial_{[i}B_{k]0})\approx 0 \\
\tfrac12 \phi_{0i}^{(3)} &= \partial_j\bigl(2\lambda_1B_{ji} - 4\lambda_2\epsilon_{0ijk}B_{0k}\bigr)
- 2\mu_1 B_{0i} - 2\lambda_1 \mu_{0i} \nonumber\\
&\quad{} + \frac{2}{\sigma_1}\lambda_2\epsilon_{0ijk}\pi_{jk}
- 4\lambda_2\epsilon_{0ijk}\partial_j B_{0k} - 2\mu_2\epsilon_{0ijk}B_{jk} \approx 0
\label{phi^3_0i}
\end{align}
By taking the combinations $\phi_{0i}^{(3)}B_{0i} - \phi_1^{(3)}\lambda_1$ and
$\phi_{0i}^{(3)}\epsilon_{0ijk}B_{jk} + 4\phi_2^{(3)}\lambda_1$ we obtain the system of equations
\begin{align}
-4B_{0i}B_{0i} \mu_1 + y_2\mu_2 + \ldots &= 0 \\
y_2\mu_1 - 8B_{jk}B_{jk} \mu_2 + \ldots &= 0
\end{align}
where we used the constraint $\phi_2^{(2)}$;
the ellipses stand for expressions that do not depend on the coefficients $\mu_1$, $\mu_2$ and $\mu_{0i}$.
If $y_2 \ne 0$ these equations can always be solved for the coefficients $\mu_1$ and $\mu_2$.
Finally, we can use constraint \eqref{phi^3_0i} to solve for the coefficient $\mu_{0i}$.
Thus, the Dirac procedure does not generate any further constraints.

Altogether there are $2\times 6 + 2\times 2 = 16$ field components (corresponding to $B_{\mu\nu}$,
the two Lagrange multipliers, and their conjugate momenta)
and 10 remaining (second-class) constraints,
so this case has 6 local degrees of freedom.
\\

Let us now see if the Hamiltonian \eqref{H_c_sigma123=0} is bounded from below.
Absence of ghost instabilities requires that $\sigma_1 < 0$.
Let us define, as in the previous case, the vectors $E_i = B_{0i}$ and
$H_i = \frac{1}{2}\epsilon_{0ijk}B_{jk}$, and moreover, $P_i = \frac{1}{2\sigma_1}\epsilon_{0ijk}\pi_{jk}$.
The secondary constraints imply conditions \eqref{constraints_AC}, as well as
\begin{align}
\vec{\nabla}\times\vec P = -\frac{1}{\sigma_1}(2\lambda_1 \vec{E} + 4\lambda_2 \vec{H})
\label{constraint_PEH}
\end{align}
The Hamiltonian density can be written as
\begin{equation}
\mathcal{H}_c \approx -\sigma_1\bigl[|\vec{P}|^2 -  2\vec{P}\cdot (\vec{\nabla}\times\vec{E})
+ (\vec{\nabla}\cdot\vec{H})^2\bigr]
\label{H_c-caseB}
\end{equation}
Now consider the configuration
\begin{align}
\vec{H} &= H\bigl(\cos\alpha \>,\> \sin\alpha \>,\> 0\bigr) \\
\vec{E} &= E\bigl(\cos(\alpha + \phi) \>,\> \sin(\alpha + \phi) \>,\> 0\bigr) \\
\vec{P} &= E\bigl(0 \>,\> 0 \>,\> \partial_x\bigl(\sin(\alpha + \phi)\bigr) - \partial_y\bigl(\cos(\alpha + \phi)\bigr)\bigr)
\end{align}
where $H = |\vec{H}|$, $E = |\vec{E}|$
and the angle $\phi$ between $\vec{H}$ and $\vec{E}$
are chosen such that
\begin{equation}
 H^2 - E^2 = -\tfrac12 y_1 \qquad\text{and}\qquad HE\cos\phi = -\tfrac12 y_2\>.
\label{constraints_case-B}
\end{equation}
We will assume that the vectors $\vec{H}$ and $\vec{E}$ are linearly independent
(i.e., $\sin\phi \ne 0$).
The function $\alpha(x,y,z)$ is taken to be of the form
\begin{equation}
\alpha(x,y,z) = \epsilon\exp[-(\kappa_x x^2 + \kappa_y y^2 + \kappa_z z^2)/2]
\end{equation} 
for positive constants $\epsilon$, $\kappa_x$, $\kappa_y$ and $\kappa_z$.
The curl of $\vec{P}$ becomes
\begin{equation}
\vec{\nabla}\times\vec{P} =
E\left(\partial_y\bigl(\cos(\alpha + \phi)\partial_x\alpha + \sin(\alpha + \phi)\partial_y\alpha\bigr) \>,\>
-\partial_x\bigl(\cos(\alpha + \phi)\partial_x\alpha + \sin(\alpha + \phi)\partial_y\alpha\bigr) \>,\> 0\right) \>.
\end{equation}
Clearly, this can be written as a linear combination of the vectors $\vec{H}$ and $\vec{E}$,
and therefore the constraint \eqref{constraint_PEH} is satisfied for suitable
values of the Lagrange multipliers $\lambda_1$ and $\lambda_2$.
A straightforward calculation yields for the Hamiltonian density \eqref{H_c-caseB}:
\begin{align}
\mathcal{H}_c &\approx -\sigma_1 \alpha^2
\bigl[\bigl(-E\cos^2(\alpha+\phi) + H^2\sin^2\alpha\bigr)\kappa_x^2 x^2
+ \bigl(-E\sin^2(\alpha+\phi) + H^2\cos^2\alpha\bigr)\kappa_y^2 y^2 \nonumber\\
&\qquad\qquad{} - \bigl(E^2\sin(2\alpha + 2\phi) - H^2\sin(2\alpha)\bigr)\kappa_x\kappa_y xy\bigr]
\end{align}
Let us now assume that $\epsilon \ll 1$.
Noting that $\alpha \le \epsilon \ll 1$,
we can expand the terms in square brackets as a power series in $\epsilon$.
For our purposes it is sufficient to just keep the leading ($\alpha$-independent) term,
and we obtain the approximate expression
\begin{equation}
\mathcal{H}_c \approx -\sigma_1\epsilon^2\bigl(-E^2\cos^2\phi \,\kappa_x^2 x^2 + (H^2 - E^2\sin^2\phi)\kappa_y^2 y^2
- E^2\sin(2\phi) \kappa_x\kappa_y xy\bigr)e^{-(\kappa_x x^2 + \kappa_y y^2 + \kappa_z z^2)}\>.
\end{equation}
This can be readily integrated over space, yielding for the energy
\begin{equation}
\int dx\,dy\,dz\,\mathcal{H}_c \approx -\sigma_1\epsilon^2\frac{\pi^{3/2}}{\sqrt{\kappa_z}}
\left(-E^2\cos^2\phi\,\sqrt{\frac{\kappa_x}{\kappa_y}}
+ (H^2 - E^2\sin^2\phi)\sqrt{\frac{\kappa_y}{\kappa_x}}\right)
\label{H_c-caseB_2}
\end{equation}
If $\cos\phi \ne 0$,
we can always choose the value of the ratio $\kappa_x/\kappa_y$ large enough such that
the expression inside square brackets is negative,
corresponding to a negative value of the total energy.
However, if $\cos\phi = 0$ (corresponding to the case $y_2 = 0$),
Eq.\ \eqref{H_c-caseB_2} reduces to
\begin{equation}
\int dx\,dy\,dz\,\mathcal{H}_c \approx -\sigma_1\epsilon^2
(H^2 - E^2)\sqrt{\frac{\pi^3\kappa_y}{\kappa_x\kappa_z}}
\end{equation}
By the first constraint in \eqref{constraints_case-B} this is negative definite for $y_1 > 0$,
but positive definite for $y_1 < 0$. 
(We will not consider the possibility $y_1 = 0$,
as this would imply there is no spontaneous Lorentz-violation.)

Lest one suspect that for $y_1 < 0$, $y_2 = 0$ the Hamiltonian might be bounded from below,
consider the configuration
\begin{align}
\vec{H} &= \bigl(H(x,y,z) \>,\> 0 \>,\> 0\bigr) \\
\vec{E} &= \bigl(0 \>,\> 0 \>,\> E(x,y,z)\bigr) \\
\vec{P} &= \bigl(0 \>,\> -\partial_x E \>,\> 0\bigr)
\end{align}
where $E(x,y,z)$ is some function of the spatial coordinates and
$H(x,y,z) = \sqrt{E(x,y,z)^2 - \frac{1}{2}y_1}$.
It follows that
\begin{equation}
\vec{\nabla}\times\vec{P} = (\partial_x\partial_z E \>,\> 0 \>,\> -\partial_x^2 E)
\end{equation}
which can be written as a linear combination of the vectors $\vec{H}$ and $\vec{E}$
for suitable values of the Lagrange multipliers $\lambda_1$ and $\lambda_2$,
satisfying constraint \eqref{constraint_PEH}.
It is easy to check that the Hamiltonian density becomes
\begin{equation}
\mathcal{H}_c = -\sigma_1\frac{E^2 - H^2}{H^2}\,(\partial_x E)^2
\end{equation}
which is negative definite for $y_1 < 0$.
\\

Let us consider now the Lagrangian \eqref{L_flat-space_lambda1}.
We have the primary constraints $\phi_1 \approx 0$ and $\phi_{0i} \approx 0$
given by expressions \eqref{phi_1_B} and \eqref{phi_0i},
so in the canonical Hamiltonian \eqref{H_c_sigma123=0},
the potential term is substituted by $V_{\lambda_1}$,
while the total Hamiltonian \eqref{H_T-caseB} the term with the parameter $\mu_2$ is set to zero. 
Applying Dirac's procedure yields the secondary constraints $\phi_1^{(j)}$ and $\phi_{0i}^{(j)}$ ($j=2,3$)
given by expressions \eqref{phi^2_1} and \eqref{phi^2_0i}--\eqref{phi^3_0i},
with $\lambda_2$ and $\mu_2$ set to zero.
The conditions $\phi_1^{(3)} = 0$ and $\phi_{0i}^{(3)} = 0$ can be solved for 
$\mu_1$ and $\mu_{0i}$, so there are no further constraints. 
There are 8 remaining constraints, which form, once more, a second-class system.
It follows that for this case there are $2\times(6 + 1) - 8 = 6$ degrees of freedom
in phase space.

Once again, in order to avoid ghost instabilities,
one needs to impose the restriction $\sigma_1 < 0$.
Defining the vectors $\vec{H}$, $\vec{E}$ and $\vec{P}$ as above,
it follows from the constraints $\phi_1^{(2)}$ and $\phi_{0i}^{(2)}$ that:
\begin{align}
\label{constraint_HE_caseB}
|\vec{H}\,|^2 - |\vec{E}\,|^2 &= -\tfrac{1}{2}y_1 \\
\vec{\nabla}\times\vec P &= -\frac{2\lambda_1}{\sigma_1}\,\vec{E}
\label{constraint_PE}
\end{align}
Now consider the ansatz
\begin{equation}
\vec{P} = \bigl(0\>,\>0\>,\>P_z(x,y,z)\bigr)
\end{equation}
with
\begin{equation}
P_z(x,y,z) = \frac{\alpha}{\bigl(1 + \kappa(x^2 + y^2 + z^2)/2\bigr)^2}\>.
\end{equation}
with positive constants $\alpha$ and $\kappa$.
If we take the Lagrange multiplier $\lambda_1$ to be given by
\begin{equation}
\lambda_1(x,y,z) = \frac{\sigma_1\sqrt{x^2 + y^2}}{\beta\bigl(1 + \kappa(x^2 + y^2 + z^2)/2\bigr)^3}
\end{equation} 
for $\beta$ constant, it follows from condition \eqref{constraint_PE} that
\begin{equation}
\vec{E} = (-y \>,\> x \>,\> 0)\frac{\alpha\beta}{\sqrt{x^2 + y^2}}\>.
\end{equation}
It is easy to check that 
\begin{equation}
\vec{\nabla}\times\vec{E} =  \bigl(0\>,\>0\>,\> \frac{\alpha\beta}{\sqrt{x^2 + y^2}}\bigr)\>.
\end{equation}
We can satisfy constraint \eqref{constraint_HE_caseB} if we choose the constants $\alpha$
and $\beta$ such that
\begin{equation}
|\vec{E}|^2 = \alpha^2\beta^2 > \tfrac12 y_1\>.
\label{condition_alpha_beta}
\end{equation}
It follows that $|\vec{H}| = ~\sqrt{\alpha^2\beta^2 - \tfrac12 y_1}\>$ is constant,
and we are free to choose $\vec{H}$ independent of the space coordinates.
It is then straightforward to check that the energy of the configuration becomes
\begin{align}
\int d^3r\,\mathcal{H}_c &= \int d^3r\,\bigl(|\vec{P}|^2 - 2\vec{P}\cdot(\vec{\nabla}\times\vec{E}) + |\vec{\nabla}\cdot\vec{H}|^2\bigr) \nonumber\\
&= \pi^2\alpha^2\left(\frac{1}{2\sqrt{2}\kappa^{3/2}} - \frac{4\beta}{\kappa}\right)
\end{align}
which is negative if $\beta > 1/(8\sqrt{2\kappa})$.
This condition is compatible with \eqref{condition_alpha_beta}.
We conclude that also in this case the energy can take arbitrary negative values.
\\

In conclusion, for the case $\sigma_{123} = 0$ the Hamiltonian is always unbounded from below.

\subsection{Case $\sigma_1 = 0$, $\sigma_{123} \ne 0$}
\label{subsec:sigma1=0}

In this case we find for the momentum conjugate to the $B_{0i}$ components
\begin{equation}
\pi_{0i} = \frac{\partial\mathcal{L}_{\lambda_1\lambda_2}}{\partial B_{0i}} = \sigma_{23}(\dot{B}_{0i} + \partial_j B_{ij})
\end{equation}
while the remaining canonical momenta vanish identically,
leading to the primary constraints
\begin{align}
\phi_1 &= \pi_{\lambda_1} \approx 0 \\
\phi_2 &= \pi_{\lambda_2} \approx 0 \\
\phi_{ij} &= \pi_{ij} \approx 0 \qquad (i,j = 1,2,3)\>.
\end{align}
%
For the canonical Hamiltonian we find
\begin{equation}
\mathcal{H}_c = \frac{1}{2\sigma_{23}}\pi_{0i}\pi_{0i} - \pi_{0i}\partial_j B_{ij}
+ \frac{\sigma_{23}}{2}\bigl(\partial_j B_{0j}\bigr)^2 + V_{\lambda_1\lambda_2}
\label{H_c_sigma1=0}
\end{equation}
and for the total Hamiltonian
\begin{equation}
\mathcal{H}_T = \mathcal{H}_c + \mu_1 \pi_{\lambda_1} + \mu_2 \pi_{\lambda_2} + \mu_{ij}\pi_{ij}
\end{equation}
where the coefficient functions $\mu_1$, $\mu_2$ and $\mu_{ij}$ are to be determined.

Just like in the previous case, we again obtain secondary constraints
demanding that time evolution preserves the primary constraints.
This yields the secondary constraints \eqref{phi^2_1} and \eqref{phi^2_2}, together with
\begin{equation}
\phi_{ij}^{(2)} = -\partial_{[i}\pi_{j]0} - 2\lambda_1 B_{ij} + 4\lambda_2\epsilon_{0ijk}B_{0k} \approx 0
\end{equation}
Imposing that the secondary constraints be conserved in time gives the (tertiary) constraints
\begin{align}
\phi_1^{(3)} &= -\frac{4}{\sigma_{23}}B_{0i}\pi_{0i} + 4B_{0i}\partial_j B_{ij}
+ 2\mu_{ij}B_{ij} \approx 0\\
\phi_2^{(3)} &= -\frac{4}{\sigma_{23}}\epsilon_{0ijk}B_{jk}\pi_{0i}
+ 4\epsilon_{0ijk}B_{jk}\partial_l B_{il}
- 4\epsilon_{0ijk}B_{0i}\mu_{jk} \approx 0 \\
\tfrac12 \phi_{ij}^{(3)} &= 2\partial_{[i}(\lambda_1 B_{0j]})
+ 2\partial_{[i}(\epsilon_{0j]kl}\lambda_2 B_{kl}) + B_{ij}\mu_1 
- \lambda_1\mu_{ij} - 2\epsilon_{0ijk}B_{0k}\mu_2
+ \frac{2}{\sigma_{23}}\lambda_2\epsilon_{0ijk}\pi_{0k} \approx 0
\label{phi^3_ij_caseC}
\end{align}
By taking the combinations $B_{ij}\phi^{(3)}_{ij} + \lambda_1\phi_1^{(3)}$
and $2\epsilon_{0ijk}B_{0i}\phi_{jk}^{(3)} - \lambda_1\phi_2^{(3)}$
we obtain the system of equations
\begin{align}
2B_{ij}B_{ij} \mu_1 - y_2\mu_2 + \ldots &= 0 \\
y_2\mu_1 - 16B_{0i}B_{0i} \mu_2 + \ldots &= 0
\end{align}
where we used the constraint $\phi_2^{(2)}$;
the ellipses stand for expressions that do not depend on the coefficients $\mu_1$, $\mu_2$ and $\mu_{ij}$.
If $y_2 \ne 0$ these equations can always be solved for the coefficients $\mu_1$ and $\mu_2$.
Finally, we can use constraint \eqref{phi^3_ij_caseC} to solve for the coefficients
$\mu_{ij}$.
Thus, the Dirac procedure does not generate any further constraints.

Just like in the previous case,
there are 16 field components and 10 remaining second-class constraints,
yielding 6 local degrees of freedom in phase space.
\\

Let us now see if the Hamiltonian \eqref{H_c_sigma1=0} is bounded from below for $\sigma_{23} > 0$
(as we saw, for negative $\sigma_{23}$ it is unbounded from below even in the absence of the constraints).
Let us define, as in the previous case, the vectors $E_i = B_{0i}$ and
$H_i = \frac{1}{2}\epsilon_{0ijk}B_{jk}$, and moreover, $P_i = \frac{1}{\sigma_{23}}\pi_{0i}$.
The secondary constraints imply conditions \eqref{constraints_AC}, as well as
\begin{align}
\vec{\nabla}\times\vec P = \frac{1}{2\sigma_{23}}(2\lambda_1 \vec{H} - 4\lambda_2 \vec{E})
\label{constraint_PEH_caseC}
\end{align}
The Hamiltonian density can then be written as
\begin{equation}
\mathcal{H}_c \approx \frac{\sigma_{23}}{2}\bigl[|\vec{P}|^2 -  2\vec{P}\cdot (\vec{\nabla}\times\vec{H})
+ (\vec{\nabla}\cdot\vec{E})^2\bigr]
\label{H_c-caseC}
\end{equation}
Comparing Eqs.\ \eqref{constraint_PEH_caseC} and \eqref{H_c-caseC} with
the corresponding Eqs.\ \eqref{constraint_PEH} and \eqref{H_c-caseB} of the previous case,
we see that they can be obtained by performing the substitutions:
\begin{equation}
\sigma_1 \to -\tfrac12 \sigma_{23}, \qquad 
\vec{E} \to \vec{H}, \qquad
\vec{H} \to -\vec{E}\>.
\label{duality}
\end{equation}
In fact, the two models are formally equivalent under these substitutions.
This is so because the Lagrangian
\eqref{L_flat-space_lambda1-lambda2} is invariant if one performs the simultaneous set of duality transformations
\begin{align}
B_{\mu\nu} &\to \frac{1}{2}\epsilon_{\mu\nu\lambda\rho}B^{\lambda\rho} \nonumber\\
\sigma_1 &\to -\sigma_1 - \tfrac12 \sigma_{23} \nonumber\\
\sigma_{23} &\to \sigma_{23} \nonumber\\
\lambda_i &\to -\lambda_i\qquad (i = 1,2)  \nonumber\\
y_i &\to -y_i\qquad (i = 1,2)
\end{align}
This transformation interchanges the cases B and C.
Thus the special case C is dynamically equivalent to the case B with Lagrangian \eqref{L_flat-space_lambda1-lambda2}.

This dynamical equivalence between the cases B and C also holds for the Lagrangian \eqref{L_flat-space_lambda1}.

Now we had already concluded that for case B the Hamiltonian is unbounded from below.
The dynamical equivalence implies that this conclusion also applies to case C. 
\\

This completes the Dirac constraint analysis of the Lagrangians
\eqref{L_flat-space_lambda1-lambda2} and \eqref{L_flat-space_lambda1}.
Our conclusion is that, for any choice of the coefficients $\sigma_i$,
the Hamiltonian is unbounded from below.
In the next section we try to remedy this by presenting a model involving the
antisymmetric two-tensor field that does have a Hamiltonian that is bounded from below,
by coupling it with the sigma-model bumblebee presented in section \ref{sec:bumblebee}.

\section{A coupled hybrid model}
\label{sec:hybrid-model}

Consider the following model,
coupling the antisymmetric two-tensor field with the sigma-model bumblebee:
\begin{equation}
\mathcal{L}_{A,B} = \mathcal{L}_A + \mathcal{L}_B\>,
\label{L-hybrid}
\end{equation}
where
\begin{equation}
\mathcal{L}_A = -\tfrac{1}{2} \bigl(\partial_\mu A_\nu\bigr)\bigl(\partial^\mu A^\nu\bigr) + \lambda_1(A^\mu A_\mu + a^2)
\label{L_A}
\end{equation}
and
\begin{equation}
\mathcal{L}_B = - \frac{\sigma_1}{2a^2}A^\mu A^\nu\bigl(\partial_\lambda B_{\mu\alpha}\bigr)\bigl(\partial^\lambda B_\nu{}^\alpha\bigr)
+ \lambda_2\left(\frac{A^\mu B_{\mu\alpha} A^\nu B_\nu{}^\alpha}{a^2} - b^2\right)
\label{L_B}
\end{equation}
with $\sigma_1$ a dimensionless constant.
The two constraints ensure that $A_\mu$ is a timelike vector satisfying
$A_\mu A^\mu = -a^2$,
while $C_\mu = A^\lambda B_{\lambda\mu}$ is spacelike,
satisfying $C_\mu C^\mu = a^2b^2$.
We already know that the first two terms yield a positive definite contribution
$\mathcal{H}_A$ to the Hamiltonian density.
The contribution of the terms involving the antisymmetric tensor field is
equal to
\begin{equation}
\mathcal{H}_B = \frac{\sigma_1}{2a^2} A^\mu A_\nu\left[\dot{B}_{\mu\alpha}\dot{B}^{\nu\alpha}
+ \bigl(\partial_i B_{\mu\alpha}\bigr)\bigl(\partial_i B^{\nu\alpha}\bigr)\right]
\label{H_B}
\end{equation}
Now let us parametrize
\begin{align}
A_0 &= a\cosh\chi\\
A_i &= a f_i\sinh\chi
\end{align}
with $f_i f_i = 1$.
Let us denote the term of Eq.\ \eqref{H_B} with time derivatives by $\mathcal{H}_{B,t}$.
Defining
\begin{equation}
D_\alpha = \frac{1}{a} A^\mu \dot{B}_{\mu\alpha}
\end{equation}
it follows that
\begin{align}
2\sigma_1^{-1}\mathcal{H}_{B,t}&= D_\alpha D^\alpha = D_k D_k - D_0^2 \nonumber\\
& = (\cosh\chi \dot{B}_{0k} - \sinh\chi f_i \dot{B}_{ik})(\cosh\chi \dot{B}_{0k} - \sinh\chi f_j \dot{B}_{jk})
- (\sinh\chi)^2(f_i \dot{B}_{i0})^2
\label{H-time-derivatives}
\end{align}
To see that this expression is positive semi-definite, note that
\begin{equation}
D_k D_k = (f_l f_l)(D_k D_k)\geqslant (f_k D_k)^2 = (\cosh\chi\,f_k \dot{B}_{0k})^2
\label{DkDk}
\end{equation}
where we used Schwarz's inequality, and thus
\begin{equation}
2\sigma_1^{-1}\mathcal{H}_{B,t} \geqslant (\cosh\chi)^2(f_k\dot{B}_{0k})^2 
- (\sinh\chi)^2(f_i \dot{B}_{i0})^2 = (f_k \dot{B}_{0k})^2 \geqslant 0\>.
\end{equation}
Note that $\sigma_1^{-1}\mathcal{H}_{B,t}$ is positive semi-definite rather than positive
definite, because there are nontrivial field configurations for which it vanishes.
More on this below.

The same logic follows with $\dot{B}_{\mu\alpha}$ replaced by
$\partial_i B_{\mu\alpha}$ for $i = 1$, 2, 3.
In conclusion, the Hamiltonian density corresponding to the
Lagrangian \eqref{L-hybrid} is positive semi-definite for $\sigma_1 > 0$.
\\

The equations of motion that follow from the Lagrangian \eqref{L-hybrid}
are
\begin{align}
\label{eq-A}
\partial^2 A_\mu + 2\lambda_1 A_\mu - \frac{\sigma_1}{a^2}A^\nu\bigl(\partial_\lambda B_{\mu\alpha}\bigr)
\bigl(\partial^\lambda B_\nu{}^\alpha\bigr) + \frac{2}{a^2}\lambda_2 B_{\mu\alpha} A^\nu B_\nu{}^\alpha &= 0 \\
\label{eq-B}
\sigma_1\partial^\lambda\bigl(A^\nu A^{[\mu}(\partial_\lambda B_\nu{}^{\rho]})\bigr)
+ 2\lambda_2 A^\nu A^{[\mu} B_\nu{}^{\rho]} &= 0\\
\label{constraint1}
A^\mu A_ \mu + a^2 &= 0\\
\label{constraint2}
A^\mu B_{\mu\alpha} A^\nu B_\nu{}^\alpha - a^2b^2 &= 0\>.
\end{align}
Contracting Eq.\ \eqref{eq-A} with $A^\mu$ and Eq.\ \eqref{eq-B} with $B_{\mu\rho}$
and using the constraints \eqref{constraint1} and \eqref{constraint2} yields
expressions for the Lagrange multipliers in terms of the fields.
Substituting these back into Eqs.\ \eqref{eq-A} and \eqref{eq-B} yields,
after some algebra, the factored equations
\begin{equation}
\bigl(\delta^\kappa_\mu + \frac{1}{a^2}A_\mu A^\kappa\bigr) \left[\partial^2 A_\kappa
- \frac{\sigma_1}{a^2}A^\nu(\partial_\lambda B_{\kappa\alpha})(\partial^\lambda B_\nu{}^\alpha)
- \frac{\sigma_1}{a^4b^2}B_{\alpha\beta}\partial_\lambda\bigl(A^\alpha A^\gamma(\partial^\lambda B_\gamma{}^\beta)\bigr)B_{\kappa\rho}A^\nu B_\nu{}^\rho\right] = 0
\label{factored-eq-A}
\end{equation}
and
\begin{equation}
\bigl(\delta^{[\mu}_{\,\vphantom{\beta}\alpha}\delta^{\rho]}_\beta
- \frac{1}{a^2b^2}A^\nu A^{[\mu}B_\nu{}^{\rho]}B_{\alpha\beta}\bigr)
\,\partial_\lambda\bigl(A^\alpha A^\gamma(\partial^\lambda B_\gamma{}^\beta)\bigr) = 0\>,
\label{factored-eq-B}
\end{equation}
where the prefactors are projectors.
Eqs.\ \eqref{factored-eq-A} and \eqref{factored-eq-B} represent massless fields,
with interactions due to the nonlinear coupling between $A_\mu$ and $B_{\mu\nu}$.
We also see clearly from Eq.\ \eqref{factored-eq-A} that $A_\mu$ continues to have 3 degrees of freedom,
while the prefactor in Eq.\ \eqref{factored-eq-B} projects out one degree of freedom,
leaving a total of $3 - 1 = 2$ propagating degrees of freedom for $B_{\mu\nu}$.

The situation is more clearly understood in the Lorentz frame in which
$A_\mu$ is purely timelike, that is, $A_0 = a$ and $A_i = 0$ ($i=1,2,3$).
Then the second constraint yields $B_{0i}B_{0i} = b^2$,
thus fixing the modulus of the three-vector $B_{0i}$.
Note that there is no constraint in the remaining components $B_{ij}$.
In this observer frame the two propagating degrees of freedom referred to above
correspond to the the fluctuations of $B_{0i}$ that are orthogonal
to its expectation value.
Clearly the three components $B_{ij}$ do not participate in the dynamics described
by the Lagrangian \eqref{L-hybrid}.

However, it is easy to extend the model and involve these components in the dynamics
as well.
To this effect, consider instead of \eqref{L-hybrid} the extended model
\begin{equation}
\mathcal{L}_\lambda = \mathcal{L}_{A,B} + \mathcal{L}_{B^*}\>,
\label{L-hybrid-extended}
\end{equation}
where
\begin{equation}
\mathcal{L}_{B^*} = - \frac{\sigma_2}{2a^2}A^\mu A^\nu\bigl(\partial_\lambda B^*_{\mu\alpha}\bigr)\bigl(\partial^\lambda B^*_\nu{}^\alpha\bigr)
+ \lambda_3\left(\frac{A^\mu B^*_{\mu\alpha} A^\nu B^*_\nu{}^\alpha}{a^2} - \bar{b}^2\right)\>,
\label{L_B^*}
\end{equation}
in which 
\begin{equation}
B^*_{\mu\nu} = \tfrac{1}{2} \epsilon_{\mu\nu\rho\sigma}B^{\rho\sigma}
\end{equation}
denotes the dual to $B_{\mu\nu}$,
and $\sigma_2$ is a dimensionless constant.
The constraint enforced by $\lambda_3$ will force an expectation value for
the components $B^*_{0i}$, or, equivalently, for the components $B_{ij}$.
It is straightforward to show that Eq.\ \eqref{L_B^*} can be written,
by using the constraints enforced by $\lambda_1$ and $\lambda_2$, as
\begin{equation}
\mathcal{L}_{B^*} = - \frac{\sigma_2}{2a^2}A^\mu A^\nu\bigl(\partial_\lambda B_{\mu\alpha}\bigr)\bigl(\partial^\lambda B_\nu{}^\alpha\bigr)
- \frac{\sigma_2}{4}\bigl(\partial_\lambda B^{\mu\nu}\bigr)\bigl(\partial^\lambda B_{\mu\nu}\bigr)
+ \lambda_3(\tfrac{1}{2} B_{\mu\nu} B^{\mu\nu} + b^2 - \bar{b}^2)\>.
\label{L_B^*-alt}
\end{equation}
Note that the first kinetic term in Eq.\ \eqref{L_B^*-alt} is identical
to the one appearing in Lagrangian \eqref{L_B},
while the second one corresponds to the ``sigma-model'' term proportional
to $\sigma_1$ considered before in Eq.\ \eqref{L_B-general}.
It follows that
\begin{equation}
\mathcal{L}_\lambda = \mathcal{L} + \lambda_1(A^\mu A_\mu + a^2)
+ \lambda_2\left(\frac{A^\mu B_{\mu\alpha} A^\nu B_\nu{}^\alpha}{a^2} - b^2\right)
+ \lambda_3\bigl(\tfrac{1}{2}B_{\mu\nu} B^{\mu\nu} + b^2 - \bar{b}^2\bigr)
\label{L-hybrid-extended2}
\end{equation}
with
\begin{equation}
\mathcal{L} = -\tfrac{1}{2} \bigl(\partial_\mu A_\nu\bigr)\bigl(\partial^\mu A^\nu\bigr) 
- \frac{\sigma_1 + \sigma_2}{2a^2} A^\mu A^\nu\bigl(\partial_\lambda B_{\mu\alpha}\bigr)\bigl(\partial^\lambda B_\nu{}^\alpha\bigr)
- \frac{\sigma_2}{4}\bigl(\partial_\lambda B^{\mu\nu}\bigr)\bigl(\partial^\lambda B_{\mu\nu}\bigr)
\end{equation}
This model has a total of $3 + 2 + 2 = 7$ propagating degrees of freedom.
Because both the original model \eqref{L-hybrid} and the added
model \eqref{L_B^*} are stable for positive values of $\sigma_1$ and $\sigma_2$,
the same is true for the extended model \eqref{L-hybrid-extended2}.

Above we argued that it was natural to extend the original model \eqref{L-hybrid}
because not all field components participate in its dynamics.
For the extended model the sum of the number of propagating degrees of freedom and the number
of constraints, $7 + 3 = 10$,
exactly matches the total number of field components ($4 + 6$).
This demonstrates that in the extended model all field components participate
in the dynamics, either as a propagating degree of freedom or because its dynamics
is fixed by the constraints.
Consequently, the Hamiltonian of the extended model is positive definite for positive
values of $\sigma_1$ and $\sigma_2$, rather than positive semi-definite.

In the limit $\sigma_2 \to 0$, while maintaining $\sigma_1 > 0$,
the Lagrangian \eqref{L-hybrid-extended2} reduces to the original model
\eqref{L-hybrid} with 5 degrees of freedom.
The case $\sigma_1 \to 0, \sigma_2 > 0$ can be
obtained from this case under the substitution $B_{\mu \nu} \to
B^*_{\mu \nu}$, and so also has 5 degrees of freedom.
Whenever either $\sigma_1$ or $\sigma_2$ is negative,
the Hamiltonian is unbounded from below.
\\

The equations of motion that result from the Lagrangian \eqref{L-hybrid-extended2} are:
\begin{align}
\label{eom-lambda-Amu}
\frac{\delta\mathcal{L}_\lambda}{\delta A^\mu} &= \frac{\delta\mathcal{L}}{\delta A^\mu}
+ 2\lambda_1 A_\mu + \frac{2}{a^2}\lambda_2 B_{\mu\rho}A^\nu B_\nu{}^\rho = 0 \\
\frac{\delta\mathcal{L}_\lambda}{\delta B_{\mu\rho}} &= \frac{\delta\mathcal{L}}{\delta B_{\mu\rho}}
+ \frac{2}{a^2}\lambda_2 A^\nu A^{[\mu} B_\nu{}^{\rho]} + \lambda_3 B^{\mu\rho} = 0
\label{eom-lambda-Bmurho}
\end{align}
with
\begin{align}
\label{eom-Amu}
\frac{\delta\mathcal{L}}{\delta A^\mu} &= \partial^2 A_\mu -\frac{\sigma_1 + \sigma_2}{a^2}A^\nu(\partial_\lambda B_{\mu\rho})(\partial^\lambda B_\nu{}^\rho)\\
\frac{\delta\mathcal{L}}{\delta B_{\mu\rho}} &= \frac{\sigma_2}{2}\partial^2 B^{\mu\rho}
+ \frac{\sigma_1 + \sigma_2}{a^2}\partial^\lambda\bigl(A^\nu A^{[\mu}\partial_\lambda B_\nu{}^{\rho]}\bigr)\>,
\label{eom-Bmurho}
\end{align}
together with the constraints
\begin{align}
\label{constraint-1}
A_\mu A^\mu + a^2 &= 0 \\
\label{constraint-2}
A^\mu B_{\mu\rho}A^\nu B_\nu{}^\rho - a^2 b^2 &= 0 \\
B_{\mu\rho}B^{\mu\rho} + 2(b^2 - \bar{b}^2) &= 0\>.
\label{constraint-3}
\end{align}
By contracting Eq.\ \eqref{eom-lambda-Amu} with $A^\mu$ and with $B^{\mu\rho}A^\sigma B_{\sigma\rho}$
and Eq.\ \eqref{eom-lambda-Bmurho} with $B_{\mu\rho}$ and with $A_\mu A^\kappa B_{\kappa\rho}$
and using the constraints \eqref{constraint-1}--\eqref{constraint-3},
one obtains a set of linear equations for the $\lambda_i$:
\begin{align}
\label{lambda-eq1}
a^2 \lambda_1 - b^2 \lambda_2 &= \frac{1}{2}A^\mu \frac{\delta\mathcal{L}}{\delta A^\mu}\\
b^2 \lambda_1 + \frac{F(A,B)}{a^4} \lambda_2 &= -\frac{1}{2a^2}B^{\mu\rho}C_\rho \frac{\delta\mathcal{L}}{\delta A^\mu}\\
b^2 \lambda_2 + (\bar{b}^2 - b^2)\lambda_3 &= -\frac{1}{2}B_{\mu\rho}\frac{\delta\mathcal{L}}{\delta B_{\mu\rho}}\\
b^2 \lambda_2 - b^2\lambda_3 &= \frac{1}{a^2}A_\mu C_\rho\frac{\delta\mathcal{L}}{\delta B_{\mu\rho}}\>.
\label{lambda-eq4}
\end{align}
Here we defined $C_\rho = A^\alpha B_{\alpha\rho}$ and
$F(A,B) = A^\mu B_{\mu\alpha}B^{\alpha\beta}B_{\beta\gamma}B^{\gamma\nu}A_\nu$
(which is not fixed by the constraints).
From Eqs.\ \eqref{lambda-eq1}--\eqref{lambda-eq4} we obtain:
\begin{align}
\label{lambda1}
\lambda_1 
&= \frac{1}{2a^2}A^\mu\frac{\delta\mathcal{L}}{\delta A^\mu} - \frac{1}{2a^2\bar{b}^2}
\left(b^2 B_{\mu\rho} + \frac{2}{a^2} (b^2 - \bar{b}^2)A_\mu C_\rho\right)\frac{\delta\mathcal{L}}{\delta B_{\mu\rho}} \\
\label{lambda2-exp1}
\lambda_2 &= -\frac{1}{2\bar{b}^2}\left(B_{\mu\rho} + \frac{2}{a^2}\left(1 - \frac{\bar{b}^2}{b^2}\right)
A_\mu C_\rho\right)\frac{\delta\mathcal{L}}{\delta B_{\mu\rho}}\\
\label{lambda2-exp2}
&= -\frac{1}{2}\frac{a^2}{a^2b^4 + F(A,B)}\bigl(b^2 A^\mu + B^{\mu\rho}C_\rho\bigr)\frac{\delta\mathcal{L}}{\delta A^\mu}\\
\label{lambda3}
\lambda_3 &= -\frac{1}{2\bar{b}^2}\left(B_{\mu\rho} + \frac{2}{a^2}A_\mu C_\rho\right)
\frac{\delta\mathcal{L}}{\delta B_{\mu\rho}}\>.
\end{align}
We see from expressions \eqref{lambda2-exp1} and \eqref{lambda2-exp2} that there are multiple ways to express
$\lambda_2$ in terms of $A^\mu$ and $B_{\mu\rho}$.
Technically, this is a consequence of the fact that the system of
equations \eqref{lambda-eq1}--\eqref{lambda-eq4} is overconstrained.
The implied identity is a nontrivial relation between the variations $\delta\mathcal{L}/\delta A^\mu$
and $\delta\mathcal{L}/\delta B_{\mu\rho}$ that is valid on the surface defined by the constraints.
Also $\lambda_1$ and $\lambda_3$ can be re-written by using this identity.

The Lagrange-multiplier terms in the equations of motion \eqref{eom-lambda-Amu} and \eqref{eom-lambda-Bmurho}
impose the constraints \eqref{constraint-1}--\eqref{constraint-3},
constraining the possible field variations accordingly.
Therefore, one expects that it should be possible to write
the resulting equations of motion in projected form. 
In Appendix \ref{app} it is shown that this is indeed the case.

\section{The linearized equations of motion}
\label{sec:linearized-eom}
Next, let us analyze the equations of motion at the linearized level,
by expressing the fields as a sum of their expectation values and fluctuations around these:
\begin{equation}
A^\mu = \bar{A}^\mu + \delta A^\mu\>, \qquad
B_{\mu\rho} = \bar{B}_{\mu\rho} + \delta B_{\mu\rho}\>.
\end{equation}
We will assume that $\bar{A}^\mu$ and $\bar{B}_{\mu\rho}$ are independent
of spacetime.
From constraint \eqref{constraint-1} it follows that we can express
\begin{equation}
\bar{A}^\mu = a\,u^\mu \qquad \text{with } u^\mu u_\mu = -1
\end{equation}
while the fluctuation vector $a^\mu$ satisfies
\begin{equation}
a^\mu u_\mu = 0\>.
\label{constraint-amu}
\end{equation}
From constraint \eqref{constraint-2} it follows that
\begin{equation}
u^\rho \bar{B}_{\rho\mu} u_\sigma \bar{B}^{\sigma\mu} = b^2
\label{constraint-2b}
\end{equation}
and 
\begin{equation}
\delta A^\rho \bar{B}_{\rho\mu}\bar{B}^{\sigma\mu}u_\sigma +
a\,u^\rho \bar{B}_{\rho\mu} u_\sigma \delta B^{\sigma\mu} = 0\>.
\label{constraint-2c}
\end{equation}
Finally, constraint \eqref{constraint-3} yields the conditions
\begin{equation}
\bar{B}^{\mu\nu}\bar{B}_{\mu\nu} = 2(\bar{b}^2 - b^2)
\label{constraint-3b}
\end{equation}
and
\begin{equation}
\bar{B}^{\mu\nu} \delta B_{\mu\nu} = 0\>.
\label{constraint-3c}
\end{equation}
It is helpful to choose an observer frame such that $u^\mu$ is purely timelike,
with $u^0 = 1$ and $u^i = 0$ ($i = 1,2,3$). 
It then follows from \eqref{constraint-amu} that $a^0 = 0$.
Eqs.\ \eqref{constraint-2b}--\eqref{constraint-3c} then yield the conditions
\begin{align}
\label{bar-B0i}
\bar{B}_{0i}\bar{B}_{0i} &= b^2 \\
\bar{B}_{ij}\bar{B}_{ij} &= 2\bar{b}^2 
\label{bar-Bij}
\end{align}
and
\begin{align}
\label{longitudinal-b0i}
\bar{B}_{0i} \delta B_{0i} &= -\frac{1}{a}\delta A^i\bar{B}_{ij}\bar{B}_{0j} \\
\bar{B}_{ij} \delta B_{ij} &= -\frac{2}{a}\delta A^i\bar{B}_{ij}\bar{B}_{0j}
\label{longitudinal-bij}
\end{align}
(note that repeated indices $i$ and $j$ are summed over).
We see clearly that there are 7 degrees of freedom:
3 spacelike components of $\delta A^\mu$,
2 (transverse) components of $\delta B_{0i}$ satisfying $\bar{B}_{0i} \delta B_{0i} = 0$
and 2 (transverse) components of $\delta B_{ij}$ satisfying $\bar{B}_{ij}\delta B_{ij} = 0$.
The respective longitudinal components of $\delta B_{0i}$ and $\delta B_{ij}$ are fixed
by Eqs.\ \eqref{longitudinal-b0i} and \eqref{longitudinal-bij} in terms of $\delta A^i$.

Linearizing the equations of motion \eqref{eom-Amu} and \eqref{eom-Bmurho}
it follows that
\begin{align}
\frac{\delta\mathcal{L}}{\delta A^\mu} &\approx \partial^2 \delta A_\mu \\
\frac{\delta\mathcal{L}}{\delta B_{\mu\rho}} &\approx 
\partial^2\left(\frac{\sigma_2}{2} \delta B^{\mu\rho} + (\sigma_1 + \sigma_2)u^\nu u^{[\mu} \delta B_\nu{}^{\rho]}\right)
\end{align}
it is straightforward to verify that the linearized form of the equations of motion
\eqref{eom-lambda-Amu} and \eqref{eom-lambda-Bmurho} with the Lagrange
multipliers given by Eqs.\ \eqref{lambda1}--\eqref{lambda3} imply that all
fluctuations are strictly massless degrees of freedom.

The degrees of freedom can be made more explicit
if we parametrize the antisymmetric tensor field as
\begin{equation}
B_{\mu\nu} =
\begin{pmatrix}
0 & E_1 & E_2 & E_3\\
-E_1 & 0 & H_3 & -H_2\\
-E_2 & -H_3 & 0 & H_1\\
-E_3 & H_2 & -H_1 & 0
\end{pmatrix}
\end{equation}
with $E_i = B_{0i}$ and $H_i = \frac{1}{2}\epsilon_{ijk}B_{jk}$.
Expressing $\vec{E} = (E_1,E_2,E_3)$ and $\vec{H} = (H_1,H_2,H_3)$ as a sum of expectation value and fluctuations:
\begin{equation}
\vec{E} = \bar{\vec{E}} + \delta\vec{E}\>, \qquad
\vec{H} = \bar{\vec{H}} + \delta\vec{H}\>,
\end{equation}
the constraints \eqref{bar-B0i} and \eqref{bar-Bij} imply that
\begin{equation}
|\bar{\vec{E}}\,| = b
\qquad\text{and}\qquad
|\bar{\vec{H}}\,| = \bar{b}\>,
\end{equation}
while the conditions \eqref{longitudinal-b0i} and \eqref{longitudinal-bij} become
\begin{equation}
\bar{\vec{E}}\cdot \delta\vec{E} = \bar{\vec{H}}\cdot \delta\vec{H}
= \frac{1}{a}(\bar{\vec{E}}\times\bar{\vec{H}})\cdot\delta\vec{A}\>.
\label{longitudinal-conditions}
\end{equation}
We see that the components of $\delta\vec{E}$ and $\delta\vec{H}$
parallel to their expectation values are fixed in terms of the component
of $\delta\vec{A}$ perpendicular to the plane defined by $\bar{\vec{E}}$ and $\bar{\vec{H}}$.
 
Through subsequent observer rotations it is always possible,
in the frame in which $u^\mu$ is purely timelike,
to transform some of the components of $\bar{\vec{E}}$ and $\bar{\vec{H}}$ to zero.
For instance, a convenient possible form is $\bar{\vec{E}} = (b,0,0)$
and $\bar{\vec{H}} = (\bar{b}\cos\theta,\bar{b}\sin\theta,0)$.
In that case the conditions \eqref{longitudinal-conditions} become
\begin{align}
\delta E_1 &= \frac{\bar{b}}{a}\delta A_3\sin\theta\\
\delta H_1\cos\theta + \delta H_2\sin\theta &= \frac{b}{a}\delta A_3\sin\theta\>.
\end{align}
A further simplification for the form of $B_{\mu\nu}$,
such as obtained in Ref.\ \cite{Altschul2009}, is not possible in our case,
as we do not have the freedom anymore to apply an observer boost
(which would affect the purely timelike form of $u^\mu$).

It is interesting to compare these massless modes with the Nambu-Goldstone (NG) modes
corresponding to the broken Lorentz generators.
Parametrizing infinitesimal Lorentz transformations as
\begin{equation}
\Lambda^\mu{}_\nu = \delta^\mu_\nu + \omega^\mu{}_\nu
\end{equation}
with the generators $\omega_{\mu\nu} = -\omega_{\nu\mu}$, we have for the NG modes%
\footnote{Note that the NG modes for the antisymmetric two-tensor field
have been called phons \cite{Altschul2009}.}
\begin{equation}
(\delta_\omega A)^\mu = \omega^\mu{}_\nu\bar{A}^\nu\>, \qquad
(\delta_\omega B)_{\mu\rho} = \omega_\mu{}^\lambda \bar{B}_{\lambda\rho} + \omega_\rho{}^\lambda\bar{B}_{\mu\lambda}\>.
\label{omega-modes}
\end{equation}
Adopting the observer frame mentioned above, with
\begin{equation}
\bar{A}^\mu = \begin{pmatrix}
a \\ 0 \\ 0 \\ 0
\end{pmatrix}\>,\qquad
\bar{B}_{\mu\rho} = \begin{pmatrix}
0 & b & 0 & 0 \\
-b & 0 & 0 & -\bar{b}\sin\theta \\
0 & 0 & 0 & \bar{b}\cos\theta \\
0 & \bar{b}\sin\theta & -\bar{b}\cos\theta & 0
\end{pmatrix} \>,
\end{equation}
it is easy to check explicitly that the modes \eqref{omega-modes} satisfy the conditions \eqref{longitudinal-conditions},
as of course they should.
It turns out that all six Lorentz generators are broken, if $\sin\theta \neq 0$.
(In the special case when $\sin\theta = 0$ there are five broken generators.)
This leaves us with a puzzle, because we know there are seven propagating massless fields.
Clearly they can not all correspond to NG modes.
The reason for the discrepancy is that, as we saw above, to each of the expectation values
$\bar{\vec{E}}$ and $\bar{\vec{H}}$ there are two independent massless modes associated.
On the other hand, applying rotation generators only yields a total of three NG modes.
This is because the rotation generators are applied simultaneously, not independently,
to the vectors $\bar{\vec{E}}$ and $\bar{\vec{H}}$.
The discrepancy in the counting can be eliminated if we add another Lagrange-multiplier term
to the Lagrangian, such as
\begin{equation}
\lambda_4(\epsilon^{\mu\nu\rho\sigma}B_{\mu\nu}B_{\rho\sigma} \pm \tilde{b}^2)
\end{equation}
for some constant $\tilde{b}$.
This will fix the angle between $\vec{E}$ and $\vec{B}$ and reduce the total number of propagating
modes to six, all of which corresponding to NG modes.

Therefore we see that, in dynamical systems depending of fields with Lorentz indices,
in the presence of Lagrange-multiplier constraints that depend on Lorentz scalars built of the fields,
forcing nonzero expectation values of the fields,
there are massless modes corresponding to fluctuations compatible with the constraints.
Part of these modes (maximally six) can be identified with NG modes,
however, there may be additional independent modes which do not correspond to NG modes.
However, in that case it is always possible to add additional independent constraints such that
the propagating massless degrees of freedom all correspond to NG modes.
Note that this argument also applies for systems with smooth scalar potentials instead of
Lagrange multiplier constraints (in this case there are additional massive modes).

To see another simple example of this,
consider first just the bumblebee with the field $A^\mu$ and Lagrangian \eqref{L}.
It has three propagating massless degrees of freedom, all corresponding to NG modes.
Now add another, independent bumblebee, constructed of the field $B^\mu$.
Then there are clearly six propagating massless degrees of freedom.
However, only five of them can be NG modes.
In order to see this, fix the expectation value of $A^\mu$ to $(a,0,0,0)$.
We can choose a frame such that the expectation value of $B^\mu$ will be of the form
$(a\cosh\xi,0,0,a\sinh\xi)$ for some value $\xi$.
This means all boost generators and two of the three rotation generators are broken.
Thus one of the propagating modes is not a NG mode.
However, we can eliminate it by adding an extra constraint, fixing the value of $A_\mu B^\mu$.
Continuing this example, we can add another bumblebee, depending on another field $C^\mu$.
We then have nine propagating massless modes.
Six of them correspond to the broken Lorentz generators, three do not.
We can eliminate the latter by adding constraints fixing $A_\mu B^\mu$, $A_\mu C^\mu$
and $B_\mu C^\mu$.
And so forth.

\section{Couplings to gravity and matter}
\label{sec:couplings}

Let us now generalize the extended hybrid model \eqref{L-hybrid-extended2} to curved
space. To this effect, we replace the partial derivatives by covariant derivatives.
Moreover, there is the possibility to add explicit couplings to curvature.
The action we consider is
\begin{align}
S &= \int d^4x\sqrt{-g}\biggl[\frac{1}{16\pi G}R -\tfrac{1}{2} \bigl(\nabla_\mu A_\nu\bigr)\bigl(\nabla^\mu A^\nu\bigr)
+ \lambda_1(A^\mu A_\mu + a^2) + \alpha R_{\mu\nu}A^\mu A^\nu \nonumber\\
&\qquad\qquad\qquad{}
- \frac{\sigma_2}{4}\bigl(\nabla_\lambda B^{\mu\nu}\bigr)\bigl(\nabla^\lambda B_{\mu\nu}\bigr)
+ \lambda_3\bigl(\tfrac{1}{2}B_{\mu\nu} B^{\mu\nu} + b^2 - \bar{b}^2\bigr)\nonumber\\
&\qquad\qquad\qquad{} + \beta_1R_{\mu\nu\kappa\lambda}B^{\mu\nu}B^{\kappa\lambda}+ \beta_2 R_{\nu\lambda}B^{\mu\nu}B_\mu{}^\lambda\nonumber\\
&\qquad\qquad\qquad{} - \frac{\sigma_1 + \sigma_2}{a^2} A^\mu A^\nu\bigl(\nabla_\lambda B_{\mu\alpha}\bigr)\bigl(\nabla^\lambda B_\nu{}^\alpha\bigr)
+ \lambda_2\left(\frac{A^\mu B_{\mu\alpha} A^\nu B_\nu{}^\alpha}{a^2} - b^2\right)\nonumber\\
&\qquad\qquad\qquad{} + \gamma_1R_{\nu\lambda}A_\mu A_\kappa B^{\mu\nu}B^{\kappa\lambda}
+ \gamma_2 R^\mu{}_{\nu\kappa\lambda}A_\mu B^\nu{}_\rho A^\kappa B^{\lambda\rho}
\biggr]\nonumber\\
&= \int d^4x\sqrt{-g}\biggl[\frac{1}{16\pi G}R -\tfrac{1}{2} \bigl(\nabla_\mu A_\nu\bigr)\bigl(\nabla^\mu A^\nu\bigr)
- \frac{\sigma_2}{4}\bigl(\nabla_\lambda B^{\mu\nu}\bigr)\bigl(\nabla^\lambda B_{\mu\nu}\bigr)
\nonumber\\
&\qquad\qquad\qquad{}- \frac{\sigma_1 + \sigma_2}{a^2} A^\mu A^\nu\bigl(\nabla_\lambda B_{\mu\alpha}\bigr)\bigl(\nabla^\lambda B_\nu{}^\alpha\bigr)+ \lambda_1(A^\mu A_\mu + a^2) + \lambda_2\left(\frac{A^\mu B_{\mu\alpha} A^\nu B_\nu{}^\alpha}{a^2} - b^2\right) 
\nonumber\\
&\qquad\qquad\qquad{}+ \lambda_3\bigl(\tfrac{1}{2}B_{\mu\nu} B^{\mu\nu} + b^2 - \bar{b}^2\bigr)
+\tau^{\mu\nu\kappa\lambda}R_{\mu\nu\kappa\lambda} + \sigma^{\mu\nu}R_{\mu\nu}\biggr]
\label{S-curved-space}
\end{align}
where $\alpha$, $\beta_1$, $\beta_2$, $\gamma_1$ and $\gamma_2$ are constants,
and
\begin{align}
\label{sigma-munu}
\sigma^{\mu\nu} &= \alpha A^\mu A^\nu + \beta_2 B^{\lambda\mu}B_\lambda{}^\nu + \gamma_1 A_\lambda A_\kappa B^{\lambda\mu}B^{\kappa\nu}\\
\tau^{\mu\nu\kappa\lambda} &= \beta_1 B^{\mu\nu}B^{\kappa\lambda} + \gamma_2 A^\mu B^\nu{}_\rho A^\kappa B^{\lambda\rho}
\label{tau-munukappalambda}
\end{align}
parametrize the couplings to the Riemann and the Ricci tensor, respectively.
We have suppressed couplings of the Ricci scalar to scalar combinations of the
vector field and/or the antisymmetric two-tensor field,
because the values of the latter are all fixed by the constraints and thus nondynamical.
Note that this would not be the case for smooth potentials instead
of the Lagrange-multiplier constraints considered here.
An action similar to Eq.\ \eqref{S-curved-space} was considered in \cite{Altschul2009},
albeit with different kinetic terms and without the terms involving the vector field.

In the gravitational sector of the Standard-Model Extension it is conventional to denote the
Lorentz-violating couplings to the traceless part of the Ricci tensor, the Weyl tensor,
and the scalar curvature as the coefficient fields $s^{\mu\nu}$ (which is taken to be traceless),
$t^{\kappa\lambda\mu\nu}$ (with the symmetries of the Weyl tensor) and $u$.
For our model \eqref{S-curved-space} it follows that
\begin{align}
t^{\mu\nu\kappa\lambda} &= \frac{2}{3}\tau^{\mu\nu\kappa\lambda} - \frac{2}{3}\tau^{\mu[\kappa\lambda]\nu}
+ \tau^{\mu[\lambda} g^{\kappa]\nu} - \tau^{\nu[\lambda} g^{\kappa]\mu} + \frac{1}{3}\tau\,g^{\mu[\kappa}g^{\lambda]\nu}\\
s^{\mu\nu} &= \sigma^{\mu\nu} + 2\tau^{\mu\nu} - \frac{1}{4}g^{\mu\nu}(\sigma + 2\tau)\\
u &= -\frac{5}{6}\tau + \frac{1}{4}\sigma
\end{align}
where we defined $\tau^{\mu\nu} = \tau^{\mu\lambda\nu}{}_\lambda$, $\tau = \tau^\mu{}_\mu$
and $\sigma = \sigma^\mu{}_\mu$, and square brackets imply anti-symmetrization.
Using the expressions \eqref{sigma-munu} and \eqref{tau-munukappalambda},
together with the three constraints, we find
\begin{align}
t^{\mu\nu\kappa\lambda} &= \frac{2}{3}\bigl(\beta_1 B^{\mu\nu}B^{\kappa\lambda} + \gamma_2 A^\mu B^\nu{}_\rho A^\kappa B^{\lambda\rho} - \beta_1 B^{\mu[\kappa}B^{\lambda]\nu} - \gamma_2 A^\mu B^{[\kappa}{}_\rho A^{\lambda]}B^{\nu\rho}\bigr)\nonumber\\
&\quad{} + \bigl(\beta_1B^{\mu\alpha}B^{[\lambda}{}_\alpha + 2\gamma_2(\bar{b}^2-b^2)A^\mu A^{[\lambda}\bigl)g^{\kappa]\nu}
- \bigl(\beta_1B^{\nu\alpha}B^{[\lambda}{}_\alpha + 2\gamma_2(\bar{b}^2-b^2)A^\nu A^{[\lambda}\bigl)g^{\kappa]\mu}\nonumber\\
&\quad{} +\frac{2}{3}(\bar{b}^2 - b^2)(\beta_1 - a^2\gamma_2)g^{\mu[\kappa}g^{\lambda]\nu} \\
\label{s-munu}
s^{\mu\nu} &= (2\beta_1 + \beta_2)B^{\mu\alpha}B^\nu{}_\alpha + \gamma_1A_\kappa A_\lambda B^{\mu\kappa} B^{\mu\lambda} + \bigl(\alpha + 4\gamma_2(\bar{b}^2 - b^2)\bigr)A^\mu A^\nu \nonumber\\
&\quad{}- \frac{1}{4}g^{\mu\nu}\bigl[2(\bar{b}^2 - b^2)(2\beta_1 + \beta_2 - 2a^2 \gamma_2) + a^2(b^2\gamma_1 - \alpha)\bigr]\\
u &= (\bar{b}^2 - b^2)\left(-\frac{5}{3}(\beta_1 - a^2\gamma_2) + \frac{1}{2}\beta_2\right) + \frac{1}{4}a^2 (b^2 \gamma_1 - \alpha)
\end{align}
These expressions generalize the gravity couplings that have been obtained for
the bumblebee \cite{Bailey:2006fd} and for the antisymmetric tensor field \cite{Altschul2009}.
\\

It is also straightforward to add matter couplings $\int d^4x\sqrt{-g}\mathcal{L}_M$  to the action.
For instance, a (CPT-violating) vector coupling could take the form
\begin{equation}
\mathcal{L}_M \supset \bigl(\xi_1 A^\mu + \xi_2 A_\lambda B^{\lambda\mu}\bigr) j_{1,\mu}
\label{vector-coupling}
\end{equation}
where $j_{1,\mu}$ is a suitable matter current and $\xi_1$ and $\xi_2$ coupling constants.
For a Dirac fermion $j_{1,\mu}$ can be taken to be
$\frac{i}{2}\bar{\psi}\overleftrightarrow{\nabla}_\mu\psi$, or, alternatively,
$e_\mu{}^a\bar{\psi}\gamma_a\psi$ (where $e_\mu{}^a$ denotes a local vierbein).
This will give rise to contributions to the coefficients $e_\mu$ or $a_\mu$
of the Standard-Model Extension, respectively,
as was shown in Ref.\ \cite{Kostelecky:2010ze} for the case of the bumblebee.
It is worthwhile to point out that the current work provides, for the first time,
an explicitly stable model, not only for the case of a vector with timelike expectation value 
(namely, the coupling parametrized by $\xi_1$ in Eq.\ \eqref{vector-coupling}),
but also for the case of a spacelike expectation value (the $\xi_2$ term).

A matter coupling involving a symmetric traceless two-tensor can be taken in analogy to
the coefficient $s^{\mu\nu}$ in Eq.\ \eqref{s-munu} as
\begin{equation}
\bigl[\xi_3 A^\mu A^\nu + \xi_4 B_\lambda{}^\mu B^{\lambda\nu} + \xi_5 A_\lambda A_\kappa B^{\lambda\mu}B^{\kappa\nu}
-\tfrac{1}{4}\bigl(2(\bar{b}^2 - b^2)\xi_4 + a^2(b^2\xi_5 - \xi_3)\bigr)g^{\mu\nu}\bigr]j_{2,\mu\nu}
\end{equation}
For a Dirac fermion the current $j_{2,\mu\nu}$ can be taken to be
$\frac{i}{2}e_{\mu}{}^a\bar{\psi}\gamma_a\overleftrightarrow{\nabla}_{\nu}\psi$.
The expectation values of $A^\mu$ and $B^{\mu\nu}$ will then give rise to 
a contribution to the SME coefficient $c_{\mu\nu}$.

Finally, an obvious candidate for a coupling involving  an antisymmetric two-tensor is
\begin{equation}
\xi_6 B_{\mu\nu} j_3^{\mu\nu}\>.
\end{equation}
For a Dirac fermion we can take 
$j_3^{\mu\nu} = \frac{i}{2}e^\mu{}_a e^\nu{}_b\,\bar{\psi}[\gamma^a,\gamma^b]\psi$,
in which case the expectation value for $B_{\mu\nu}$ will lead to a contribution
to the SME coefficient $H^{\mu\nu}$ \cite{Altschul2009}.

\section{Conclusions}
\label{sec:conclusions}

In this work we investigated the stability of systems involving a rank-two
antisymmetric tensor field with one or two Lagrange multiplier terms forcing the latter to acquire
a vacuum expectation value, thereby breaking Lorentz invariance spontaneously.
In alternative,
we considered Lagrange-multiplier potentials fixing the value of both (pseudo-)scalars
$B_{\mu\nu}B^{\mu\nu}$ and $\epsilon^{\mu\nu\rho\sigma}B_{\mu\nu}B_{\rho\sigma}$,
or only the former one.
In the first part of this work
we performed an exhaustive Hamiltonian Dirac constraint analysis of Lagrangians
involving all possible linear combinations of kinetic terms
that are parity-even and quadratic in the antisymmetric tensor field.
It turned out to be necessary to consider separately two edge cases,
one of which corresponding to the (gauge-invariant) Maxwell-like kinetic term.
Surprisingly, these two cases are related through a duality symmetry,
which, to our knowledge, has not been pointed out before in the literature.
Our conclusion is that no kinetic term that is parity-even and quadratic in the field
leads to a Hamiltonian that is bounded from below.

We then proposed a hybrid model combining a vector field and an antisymmetric two-tensor field,
with a combination of three kinetic terms, one of which involving both fields. 
We showed this model does have a Hamiltonian that is bounded from below
in Minkowski space and is therefore stable.
We extended it to curved space, with a number of coupling terms to the curvature,
and worked out the corresponding couplings in pure gravity sector
of the Standard-Model Extension.
The formulation of this model extends the class of known stable models exhibiting
spontaneous Lorentz breaking,
up to now only encompassing that of a single vector field
with timelike expectation value.

We do not claim to have found the most general stable Lorentz-breaking system
describing an antisymmetric two-tensor field and a vector. 
In particular, we have not explored kinetic or potential terms that are
parity odd, and neither those that are higher than quadratic in the
antisymmetric tensor field.

It would be interesting to search for nontrivial solutions in curved space,
for instance those of of black-hole type.
Note that the hybrid model should allow for more general stable Lorentz-violating
backgrounds than the bumblebee models that have been considered in the literature
so far. 
It would also be interesting to search for cosmological solutions.
Presumably, it will be necessary to go beyond those of the
Friedman-Lemaitre-Robertson-Walker (FLRW) type,
because any nontrivial background of the antisymmetric two-tensor field will
necessarily be direction dependent.

Another issue that we have not considered is a full analysis of the dynamics of the degrees
of freedom of the curved-space hybrid model \eqref{S-curved-space}.
For the curved-space version of the sigma-model bumblebee presented in section \ref{sec:bumblebee},
such an analysis has been worked out in detail (this model corresponds
to the Einstein-\ae ther theory with $c_1 \ne 0$, $c_2 = c_3 = c_4 = 0$)
\cite{Jacobson:2004ts}.
For our hybrid model the analysis is likely to be considerably more involved.
In particular, the expectation value of the antisymmetric tensor field
necessarily breaks rotational invariance,
which means that the propagation of the gravitational, vector and
antisymmetric tensor waves will become direction dependent.
Such an analysis would also serve to investigate its consistency,
as theories of spin fields coupled to gravity are known to possess potential
instabilities \cite{Isenberg:1977fs}.

\acknowledgments
It is a pleasure to thank Alan Kosteleck\'y for discussions and helpful suggestions.
Financial support from Funda\c c\~ao para a Ci\^encia e a Tecnologia (Portugal)
through the research grant doi.org/10.54499/UIDB/00099/2020 is gratefully acknowledged.

\appendix
\section{Projective form of the equations of motion of the hybrid model}
\label{app}

The Lagrange-multiplier terms in the equations of motion \eqref{eom-lambda-Amu} and \eqref{eom-lambda-Bmurho}
project the latter on restricted subspace corresponding to the field variations
that are compatible with the constraints.

Let us first consider constraint \eqref{constraint-1}.
Field variations $\delta A^\mu$ compatible with this constraint satisfy
\begin{equation}
\delta A^\mu A_\mu = 0\>.
\end{equation}
We can implement this by replacing any arbitrary field variation $\delta A^\mu$
with $(P_1)^\mu_\nu \,\delta A^\nu$, with the projector $P_1$ defined by
\begin{equation}
(P_1)^\mu_\nu = \delta_\nu^\mu + \frac{1}{a^2}A^\mu A_\nu\>.
\end{equation}
Variations of the Lagrangian density $\mathcal{L}$ compatible with
constraint \eqref{constraint-1} can be expressed as
\begin{equation}
\delta\mathcal{L} = \frac{\delta\mathcal{L}}{\delta A^\mu}(P_1)^\mu_\nu\delta A^\nu
+ \frac{\delta\mathcal{L}}{\delta B_{\mu\rho}}\delta B_{\mu\rho}\>.
\end{equation}
Instead of taking the projector $P_1$ to act to the right on the field variation
$\delta A$,
we can choose to take it to act to the left on the equation of motion
$\delta\mathcal{L}/\delta A^\mu$, leaving the field variation arbitrary.
This is exactly what happened in the factored equation of motion we obtained
for the sigma-bumblebee model, Eq.\ \eqref{bumblebee-factored}.
This way, inclusion of the constraint has the effect of replacing the equation
of motion by its projected version:
\begin{equation}
\frac{\delta\mathcal{L}}{\delta A^\mu} \to
\left(\frac{\delta\mathcal{L}}{\delta A^\mu}\right)_\perp =
\frac{\delta\mathcal{L}}{\delta A^\nu}(P_1)^\nu_\mu
\end{equation}
This projected equation of motion identically satisfies the condition
\begin{equation}
\left(\frac{\delta\mathcal{L}}{\delta A^\mu}\right)_\perp A^\mu = 0\>.
\label{empty}
\end{equation}
In other words, Eq.\ \eqref{empty} is an empty condition that is identically satisfied.

In the case at hand the situation is rather more complicated,
as we have to satisfy three constraints \eqref{constraint-1}--\eqref{constraint-3}
rather than just one.
The strategy we will take is to implement this by applying a suitable prohector
in the space of the equations of motion.
The first step is to combine all equations of motion in
a 10-dimensional vector space  $V = \{\vec{v}\}$, with components
\begin{equation}
\vec{v} \equiv \begin{pmatrix}
\delta\mathcal{L}/\delta A^\mu\\
\delta\mathcal{L}/\delta B_{\mu\rho}
\end{pmatrix}\>.
\label{v-eom}
\end{equation}
Variations compatible with the constraints satisfy the three conditions
\begin{equation}
\delta A^\mu A_\mu = \delta A^\mu B_{\mu\rho}C^\rho + \delta B_{\mu\rho}A^\mu C^\rho 
= \delta B_{\mu\rho} B^{\mu\rho} = 0\>.
\label{constrained-variations}
\end{equation}
Defining the vectors
\begin{equation}
\vec{a}_1 = \begin{pmatrix}
A^\mu \\ 0
\end{pmatrix}\>,\qquad
\vec{a}_2 = \begin{pmatrix}
B^{\mu\rho}C_\rho \\ A_{[\mu}C_{\rho]}
\end{pmatrix}\>,\qquad
\vec{a}_3 = \begin{pmatrix}
0 \\ B_{\mu\rho}
\end{pmatrix}\>,
\end{equation}
the elements in the subspace $V_\perp = \{\vec{v}_\perp\}$ of $V$ compatible
with the conditions \eqref{constrained-variations} formally satisfy the orthonormality relations
\begin{equation}
\vec{v}_\perp \cdot \vec{a}_1 = \vec{v}_\perp \cdot \vec{a}_2 = \vec{v}_\perp \cdot \vec{a}_3 = 0\>.
\label{v-perp-conditions}
\end{equation} 
Here the inner vector product amounts to contraction of corresponding Lorentz indices.
Taking the ansatz
\begin{equation}
\vec{v}_\perp = \vec{v} - \beta_1 \vec{a}_1 - \beta_2 \vec{a}_2 - \beta_3 \vec{a}_3
\label{v-perp}
\end{equation}
and imposing the conditions \eqref{v-perp-conditions} it follows that the coefficients $\beta_i$
are given by
\begin{equation}
\beta_i = (M_a^{-1})_{ij}\,\vec{a}_j\cdot\vec{v}\qquad (1\le i,j\le 3)\>,
\label{beta-i}
\end{equation}
where the coefficient matrix $(M_a)_{ij}$ is defined by
\begin{equation}
(M_a)_{ij} = \vec{a}_i\cdot\vec{a}_j\>.
\end{equation}
We can evaluate its elements explicitly by using the constraints \eqref{constraint-1}-\eqref{constraint-3}:
\begin{align}
\vec{a}_1\cdot\vec{a}_1 &= A^\mu A_\mu = - a^2 \\
\vec{a}_1\cdot\vec{a}_2 &= A_\mu B^{\mu\rho}C_\rho = a^2 b^2\\
\vec{a}_1\cdot\vec{a}_3 &= 0\\
\vec{a}_2\cdot\vec{a}_2 &= B^{\mu\rho}C_\rho B_{\mu\sigma}C^\sigma + A^{[\mu}C^{\rho]}A_{[\mu}C_{\rho]}
= F(A,B) - \tfrac{1}{2}a^4b^2\\
\vec{a}_2\cdot\vec{a}_3 &= A^\mu C^\rho B_{\mu\rho} = a^2 b^2\\
\vec{a}_3\cdot\vec{a}_3 &= B_{\mu\rho}B^{\mu\rho} = 2(\bar{b}^2 - b^2)\>.
\end{align}
Straightforward calculation shows that the coefficients $\beta_i$ in Eq.\ \eqref{v-perp}
correspond exactly to the corresponding expressions obtained for the Lagrange multipliers
$\lambda_i$ in Eqs.\ \eqref{lambda1}--\eqref{lambda3},
where it is necessary to make repeated use of the nontrivial identity
implied by Eqs.\ \eqref{lambda2-exp1} and \eqref{lambda2-exp2}.
We therefore conclude that the equations of motion that belong to the projected vector space $V_\perp$
coincide exactly with Eqs.\ \eqref{eom-lambda-Amu} and \eqref{eom-lambda-Bmurho}
after substituting expressions \eqref{lambda1}--\eqref{lambda3} for the Lagrange multipliers.

The projective representation allows for expressing the equations of motion in factored form,
analogously to Eq.\ \eqref{bumblebee-factored}, by combining Eqs.\ \eqref{v-perp} and \eqref{beta-i}:
\begin{equation}
v_{\perp,\alpha} = \bigl[\delta_{\alpha\beta} - a_{i\alpha}(M^{-1}_a)_{ij}a_{j\beta}\bigr]v_\beta\>.
\end{equation}
Here the indices $\alpha$ and $\beta$ parametrize the (ten-dimensional) space of equations of motion.
Explicitly, this yields the expressions
\begin{align}
\left(\frac{\delta\mathcal{L}}{\delta A^\mu}\right)_\perp &=
\Bigl\{\delta_\mu^\nu - \bigl[A_\mu(M_a^{-1})_{11} + B_{\mu\rho}C^\rho(M_a^{-1})_{21}\bigr]A^\nu \nonumber\\
&\qquad\quad
{}- \bigl[A_\mu(M_a^{-1})_{12} + B_{\mu\rho}C^\rho(M_a^{-1})_{22}\bigr]B^{\nu\sigma}C_\sigma\Bigr\}
\frac{\delta\mathcal{L}}{\delta A^\nu}\nonumber\\
&\qquad
{}- \bigl[A_\mu(M_a^{-1})_{12} + B_{\mu\rho}C^\rho(M_a^{-1})_{22}\bigr]A_\nu C_\sigma
\frac{\delta\mathcal{L}}{\delta B_{\nu\sigma}}\\
\left(\frac{\delta\mathcal{L}}{\delta B_{\mu\rho}}\right)_\perp &=
\Bigl\{\delta^{[\mu}_\nu \delta^{\rho]}_\sigma - \bigl[B^{\mu\rho}(M_a^{-1})_{33} + A^{[\mu} C^{\rho]}(M_a^{-1})_{23}\bigr]B_{\nu\sigma} \nonumber\\
&\qquad\qquad\quad
{}- \bigl[B^{\mu\rho}(M_a^{-1})_{32} + A^{[\mu} C^{\rho]}(M_a^{-1})_{22}\bigr]A_\nu C_\sigma\Bigr\}
\frac{\delta\mathcal{L}}{\delta B_{\nu\sigma}}\nonumber\\
&\qquad
{}- \bigl[B^{\mu\rho}(M_a^{-1})_{32} + A^{[\mu}C^{\rho]}(M_a^{-1})_{22}\bigr]B^{\nu\sigma} C_\sigma
\frac{\delta\mathcal{L}}{\delta A^\nu}\>.
\end{align}


\begin{thebibliography}{xx}


\bibitem{Kostelecky:1988zi}
V.~A.~Kostelecky and S.~Samuel,
``Spontaneous Breaking of Lorentz Symmetry in String Theory,''
Phys.\ Rev.\ D \textbf{39}, 683 (1989).

\bibitem{Kostelecky:1989jw}
V.~A.~Kostelecky and S.~Samuel,
``Gravitational Phenomenology in Higher Dimensional Theories and Strings,''
Phys.\ Rev.\ D \textbf{40}, 1886-1903 (1989).

\bibitem{Kostelecky:1989jp}
V.~A.~Kostelecky and S.~Samuel,
``Phenomenological Gravitational Constraints on Strings and Higher Dimensional Theories,''
Phys.\ Rev.\ Lett.\ \textbf{63}, 224 (1989).



\bibitem{Kostelecky:1994rn}
V.~A.~Kostelecky and R.~Potting,
``CPT, strings, and meson factories,''
Phys.\ Rev.\ D \textbf{51}, 3923-3935 (1995)
[arXiv:hep-ph/9501341].

\bibitem{Colladay:1996iz}
D.~Colladay and V.~A.~Kostelecky,
``CPT violation and the standard model,''
Phys.\ Rev.\ D \textbf{55}, 6760-6774 (1997)
[arXiv:hep-ph/9703464].´

\bibitem{Colladay:1998fq}
D.~Colladay and V.~A.~Kostelecky,
``Lorentz violating extension of the standard model,''
Phys.\ Rev.\ D \textbf{58}, 116002 (1998).
[arXiv:hep-ph/9809521].




\bibitem{Kostelecky:2003fs}
V.~A.~Kostelecky,
``Gravity, Lorentz violation, and the standard model,''
Phys.\ Rev.\ D \textbf{69}, 105009 (2004)
[arXiv:hep-th/0312310].

\bibitem{Bluhm:2004ep}
R.~Bluhm and V.~A.~Kostelecky,
``Spontaneous Lorentz violation, Nambu-Goldstone modes, and gravity,''
Phys.\ Rev.\ D \textbf{71}, 065008 (2005)
[arXiv:hep-th/0412320].

\bibitem{Bailey:2006fd}
Q.~G.~Bailey and V.~A.~Kostelecky,
``Signals for Lorentz violation in post-Newtonian gravity,''
Phys.\ Rev.\ D \textbf{74}, 045001 (2006)
[arXiv:gr-qc/0603030].

\bibitem{Chkareuli:2007da}
J.~L.~Chkareuli, C.~D.~Froggatt, J.~G.~Jejelava and H.~B.~Nielsen,
``Constrained gauge fields from spontaneous Lorentz violation,''
Nucl.\ Phys.\ B \textbf{796}, 211-223 (2008)
[arXiv:0710.3479 [hep-th]].



\bibitem{Jacobson:2000xp}
T.~Jacobson and D.~Mattingly,
``Gravity with a dynamical preferred frame,''
Phys.\ Rev.\ D \textbf{64}, 024028 (2001)
[arXiv:gr-qc/0007031].

\bibitem{Jacobson:2004ts}
T.~Jacobson and D.~Mattingly,
``Einstein-Aether waves,''
Phys.\ Rev.\ D \textbf{70}, 024003 (2004)
[arXiv:gr-qc/0402005].

\bibitem{Deserfest}
C.~Eling, T.~Jacobson and D.~Mattingly,
``Einstein-Aether theory,''
\textit{Deserfest}, Ed.\ J.~Liu, M.~J.~Duff, K.~Stelle,
and R.~P.~Woodard (Singapore: World Scientific),
[arXiv:gr-qc/0410001].

\bibitem{Carroll:2009en}
S.~M.~Carroll, T.~R.~Dulaney, M.~I.~Gresham and H.~Tam,
``Sigma-Model Aether,''
Phys.\ Rev.\ D \textbf{79}, 065012 (2009)
[arXiv:0812.1050 [hep-th]].

\bibitem{Withers:2009qg}
B.~Withers,
``Einstein-aether as a quantum effective field theory,''
Class.\ Quant.\ Grav.\ \textbf{26}, 225009 (2009)
[arXiv:0905.2446 [gr-qc]].





\bibitem{Gripaios:2004ms}
B.~M.~Gripaios,
``Modified gravity via spontaneous symmetry breaking,''
JHEP \textbf{10}, 069 (2004)
[arXiv:hep-th/0408127].

\bibitem{Bertolami:2005bh}
O.~Bertolami and J.~Paramos,
``The Flight of the bumblebee: Vacuum solutions of a gravity model with vector-induced spontaneous Lorentz symmetry breaking,''
Phys.\ Rev.\ D \textbf{72}, 044001 (2005)
[arXiv:hep-th/0504215].

\bibitem{Graesser:2005bg}
M.~L.~Graesser, A.~Jenkins and M.~B.~Wise,
``Spontaneous Lorentz violation and the long-range gravitational preferred-frame effect,''
Phys.\ Lett.\ B \textbf{613}, 5-10 (2005)
[arXiv:hep-th/0501223].

\bibitem{Eling:2006ec}
C.~Eling and T.~Jacobson,
``Black Holes in Einstein-Aether Theory,''
Class.\ Quant.\ Grav.\ \textbf{23}, 5643-5660 (2006)
[erratum: Class.\ Quant.\ Grav.\ \textbf{27}, 049802 (2010)]
[arXiv:gr-qc/0604088].

\bibitem{Garfinkle:2007bk}
D.~Garfinkle, C.~Eling and T.~Jacobson,
``Numerical simulations of gravitational collapse in Einstein-aether theory,''
Phys.\ Rev.\ D \textbf{76}, 024003 (2007)
[arXiv:gr-qc/0703093].

\bibitem{Barausse:2015frm}
E.~Barausse, T.~P.~Sotiriou and I.~Vega,
``Slowly rotating black holes in Einstein-\ae{}ther theory,''
Phys.\ Rev.\ D \textbf{93}, no.4, 044044 (2016)
[arXiv:1512.05894 [gr-qc]].

\bibitem{Casana:2017jkc}
R.~Casana, A.~Cavalcante, F.~P.~Poulis and E.~B.~Santos,
``Exact Schwarzschild-like solution in a bumblebee gravity model,''
Phys.\ Rev.\ D \textbf{97}, no.10, 104001 (2018)
[arXiv:1711.02273 [gr-qc]].

\bibitem{Ding:2020kfr}
C.~Ding and X.~Chen,
``Slowly rotating Einstein-bumblebee black hole solution and its greybody factor in a Lorentz violation model,''
Chin.\ Phys.\ C \textbf{45}, no.2, 025106 (2021)
[arXiv:2008.10474 [gr-qc]].



\bibitem{Moffat:2002nu}
J.~W.~Moffat,
``Spontaneous violation of Lorentz invariance and ultra-high-energy cosmic rays,''
Int.\ J.\ Mod.\ Phys.\ D \textbf{12}, 1279-1288 (2003)
[arXiv:hep-th/0211167].

\bibitem{Oost:2018tcv}
J.~Oost, S.~Mukohyama and A.~Wang,
``Constraints on Einstein-aether theory after GW170817,''
Phys.\ Rev.\ D \textbf{97}, no.12, 124023 (2018)
[arXiv:1802.04303 [gr-qc]].

\bibitem{Khodadi:2020gns}
M.~Khodadi and E.~N.~Saridakis,
``Einstein-\AE{}ther gravity in the light of event horizon telescope observations of M87*,''
Phys.\ Dark Univ.\ \textbf{32}, 100835 (2021)
[arXiv:2012.05186 [gr-qc]].

\bibitem{Liang:2022hxd}
D.~Liang, R.~Xu, X.~Lu and L.~Shao,
``Polarizations of gravitational waves in the bumblebee gravity model,''
Phys.\ Rev.\ D \textbf{106}, no.12, 124019 (2022)
[arXiv:2207.14423 [gr-qc]].



\bibitem{Carroll:2004ai}
S.~M.~Carroll and E.~A.~Lim,
``Lorentz-violating vector fields slow the universe down,''
Phys.\ Rev.\ D \textbf{70}, 123525 (2004)
[arXiv:hep-th/0407149].

\bibitem{Bekenstein:2004ne}
J.~D.~Bekenstein,
``Relativistic gravitation theory for the MOND paradigm,''
Phys.\ Rev.\ D \textbf{70}, 083509 (2004)
[erratum: Phys.\ Rev.\ D \textbf{71}, 069901 (2005)]
[arXiv:astro-ph/0403694].

\bibitem{Zlosnik:2006zu}
T.~G.~Zlosnik, P.~G.~Ferreira and G.~D.~Starkman,
``Modifying gravity with the Aether: An alternative to Dark Matter,''
Phys.\ Rev.\ D \textbf{75}, 044017 (2007)
[arXiv:astro-ph/0607411].

\bibitem{Tartaglia:2007mh}
A.~Tartaglia and N.~Radicella,
``Vector field theories in cosmology,''
Phys.\ Rev.\ D \textbf{76}, 083501 (2007)
[arXiv:0708.0675 [gr-qc]].

\bibitem{Barrow:2012qy}
J.~D.~Barrow,
``Some Inflationary Einstein-Aether Cosmologies,''
Phys.\ Rev.\ D \textbf{85}, 047503 (2012)
[arXiv:1201.2882 [gr-qc]].





\bibitem{Carroll2009}
S.~M.~Carroll, T.~R.~Dulaney, M.~I.~Gresham and H.~Tam,
``Instabilities in the Aether,''
Phys.\ Rev.\ D \textbf{79}, 065011 (2009)
[arXiv:0812.1049 [hep-th]].

\bibitem{BluhmPotting}
R.~Bluhm, N.~L.~Gagne, R.~Potting and A.~Vrublevskis,
``Constraints and Stability in Vector Theories with Spontaneous Lorentz Violation,''
Phys.\ Rev.\ D \textbf{77}, 125007 (2008)
[erratum: Phys.\ Rev.\ D \textbf{79}, 029902 (2009)]
[arXiv:0802.4071 [hep-th]].

\bibitem{Clayton:2001vy}
M.~A.~Clayton,
``Causality, shocks and instabilities in vector field models of Lorentz symmetry breaking,''
[arXiv:gr-qc/0104103].

\bibitem{Elliott:2005va}
J.~W.~Elliott, G.~D.~Moore and H.~Stoica,
``Constraining the new Aether: Gravitational Cerenkov radiation,''
JHEP \textbf{08}, 066 (2005)
[arXiv:hep-ph/0505211].

\bibitem{Eling:2005zq}
C.~Eling,
``Energy in the Einstein-aether theory,''
Phys.\ Rev.\ D \textbf{73}, 084026 (2006)
[erratum: Phys.\ Rev.\ D \textbf{80}, 129905 (2009)]
[arXiv:gr-qc/0507059].

\bibitem{Seifert:2007fr}
M.~D.~Seifert,
``Stability of spherically symmetric solutions in modified theories of gravity,''
Phys.\ Rev.\ D \textbf{76}, 064002 (2007)
[arXiv:gr-qc/0703060].

\bibitem{Donnelly:2010qd}
W.~Donnelly and T.~Jacobson,
``Stability of the aether,''
Phys.\ Rev.\ D \textbf{82}, 081501(R) (2010)
[arXiv:1008.4351 [hep-th]].




\bibitem{Kostelecky:2009zr}
V.~A.~Kostelecky and R.~Potting,
``Gravity from spontaneous Lorentz violation,''
Phys.\ Rev.\ D \textbf{79}, 065018 (2009)
[arXiv:0901.0662 [gr-qc]].

\bibitem{Kraus:2002sa}
P.~Kraus and E.~T.~Tomboulis,
``Photons and gravitons as Goldstone bosons, and the cosmological constant,''
Phys.\ Rev.\ D \textbf{66}, 045015 (2002)
[arXiv:hep-th/0203221].

\bibitem{Kostelecky:2005ic}
V.~A.~Kostelecky and R.~Potting,
``Gravity from local Lorentz violation,''
Gen.\ Rel.\ Grav.\ \textbf{37}, 1675-1679 (2005)
[arXiv:gr-qc/0510124 [gr-qc]].




\bibitem{Altschul2009}
B.~Altschul, Q.~G.~Bailey and V.~A.~Kostelecky,
``Lorentz violation with an antisymmetric tensor,''
Phys.\ Rev.\ D \textbf{81} (2010), 065028
[arXiv:0912.4852 [gr-qc]].



\bibitem{Seifert:2010uu}
M.~D.~Seifert,
``A Monopole solution in a Lorentz-violating field theory,''
Phys.\ Rev.\ Lett.\ \textbf{105}, 201601 (2010)
[arXiv:1008.0324 [hep-th]].

\bibitem{Lessa:2019bgi}
L.~A.~Lessa, J.~E.~G.~Silva, R.~V.~Maluf and C.~A.~S.~Almeida,
``Modified black hole solution with a background Kalb\textendash{}Ramond field,''
Eur.\ Phys.\ J.\ C \textbf{80}, no.4, 335 (2020)
[arXiv:1911.10296 [gr-qc]].

\bibitem{Assuncao:2019azw}
J.~F.~Assun\c{c}\~ao, T.~Mariz, J.~R.~Nascimento and A.~Y.~Petrov,
``Dynamical Lorentz symmetry breaking in a tensor bumblebee model,''
Phys.\ Rev.\ D \textbf{100}, no.8, 085009 (2019)
[arXiv:1902.10592 [hep-th]].

\bibitem{Maluf:2018jwc}
R.~V.~Maluf, A.~A.~Ara\'ujo Filho, W.~T.~Cruz and C.~A.~S.~Almeida,
``Antisymmetric tensor propagator with spontaneous Lorentz violation,''
EPL \textbf{124}, no.6, 61001 (2018)
[arXiv:1810.04003 [hep-th]].

\bibitem{Aashish:2018lhv}
S.~Aashish, A.~Padhy, S.~Panda and A.~Rana,
``Inflation with an antisymmetric tensor field,''
Eur.\ Phys.\ J.\ C \textbf{78}, no.11, 887 (2018)
[arXiv:1808.04315 [gr-qc]].



\bibitem{Hernaski:2016dyk}
C.~A.~Hernaski,
``Spontaneous Breaking of Lorentz Symmetry with an antisymmetric tensor,''
Phys.\ Rev.\ D \textbf{94}, no.10, 105004 (2016)
[arXiv:1608.00829 [hep-th]].

\bibitem{Kostelecky:2010ze}
V.~A.~Kostelecky and J.~D.~Tasson,
``Matter-gravity couplings and Lorentz violation,''
Phys.\ Rev.\ D \textbf{83}, 016013 (2011)
[arXiv:1006.4106 [gr-qc]].

\bibitem{Isenberg:1977fs}
J.~A.~Isenberg and J.~M.~Nester,
``The Effect of Gravitational Interaction on Classical Fields: A Hamilton-Dirac Analysis,''
Annals Phys.\ \textbf{107}, 56-81 (1977)

\end{thebibliography}
\end{document}